 \documentclass[aps,twocolumn,pra,superscriptaddress,showpacs,tightenlines]{revtex4}
\usepackage{amsmath}
\usepackage{graphicx}
\usepackage{amssymb}
\usepackage{epsfig}
\usepackage{amsfonts}

 \def\extra#1{{}}
\def\>{\rangle}

\begin{document}
\title{Sudden vanishing of spin squeezing under decoherence}
\author{Xiaoguang Wang}
\affiliation{Advanced Science Institute, The Institute of Physical
and Chemical Research (RIKEN), Wako-shi, Saitama 351-0198, Japan}
\affiliation{Zhejiang Institute of Modern Physics, Department of
Physics, Zhejiang University, Hangzhou 310027, People's Republic
of China}
\author{Adam Miranowicz}
\affiliation{Advanced Science Institute, The Institute of Physical and Chemical Research
(RIKEN), Wako-shi, Saitama 351-0198, Japan}
\affiliation{Faculty of Physics, Adam Mickiewicz University, 61-614 Pozna\'n, Poland}
\author{Yu-xi Liu}
\affiliation{Advanced Science Institute, The Institute of Physical
and Chemical Research (RIKEN), Wako-shi, Saitama 351-0198, Japan}
\affiliation{Institute of Microelectronics and Tsinghua National
Laboratory for Information Science and Technology, Tsinghua
University, Beijing 100084, People's Republic of China}
\author{C. P. Sun}
\affiliation{Institute of Theoretical Physics, Chinese Academy of
Sciences, Beijing 100190, People's Republic of China}
\author{Franco Nori}
\affiliation{Advanced Science Institute, The Institute of Physical
and Chemical Research (RIKEN), Wako-shi, Saitama 351-0198, Japan}
\affiliation{Department of Physics, Center for Theoretical
Physics, The University of Michigan, Ann Arbor, Michigan
48109-1120, USA}

\begin{abstract}
In order to witness multipartite correlations beyond pairwise
entanglement, spin-squeezing parameters are analytically
calculated for a spin ensemble in a collective initial state under
three different decoherence channels. It is shown that, in analogy
to pairwise entanglement, the spin squeezing described by
different parameters can suddenly become zero at different
vanishing times. This finding shows the general occurrence of
sudden vanishing phenomena of quantum correlations in many-body
systems, which here is referred to as spin-squeezing sudden death
(SSSD). It is shown that the SSSD usually occurs due to
decoherence and that SSSD never occurs for some initial states in
the amplitude-damping channel. We also analytically obtain the
vanishing times of spin squeezing.
\end{abstract}
 \maketitle

\section{Introduction}

Quantum entanglement~\cite{SSS} plays an important role in both
the foundations of quantum physics and quantum-information
processing~\cite{Nielsen}. Moreover, various entangled states have
been produced in many experiments for different goals when
studying various nonclassical phenomena and their
applications~\cite{Shiyh,Polzik,Lukin,Leibfried,Guogc,Panjw,Dujf,Appel,Andre}.
Thus, entanglement is a quantum resource, and how to measure and
detect entanglement is very crucial for both theoretical
investigations and potential practical applications.

For a system of two spin-1/2 particles or a composite system of a
spin-1/2 and a spin-1, there are operationally computable
entanglement measures such as concurrence~\cite{Conc} and
negativity~\cite{Neg,Horodecki}, but no universal measures have
been found for general many-body systems. To overcome this
difficulty, entanglement witnesses are presented to detect some
kinds of entanglement in many-body
systems~\cite{Horodecki,Witness}. Now it is believed that spin
squeezing~\cite{KU, Wineland} may be useful for this
task~\cite{Sorensen,Toth,Korbicz}. In a general sense,
spin-squeezing parameters are multipartite entanglement witnesses.
For a class of many-particle states, it has been proved that the
concurrence is linearly related to some squeezing
parameters~\cite{WangSanders}. In fact, spin-squeezing
parameters~\cite{KU,Wineland,Sorensen,Toth} could be calculated
also in a simple operational fashion, which characterizes
multipartite quantum correlations beyond the pairwise
entanglement. Another important reason for choosing spin-squeezing
parameters as indicators of multipartite correlations is that spin
squeezing is relatively easy to generate~\cite{Wineland,Berman}
and measure experimentally~\cite{Fernholz,Takano}.

Besides being a parameter characterizing multipartite
correlations, spin squeezing is physically natural for controlling
many-body systems. It is difficult to control a quantum many-body
system since its constituents cannot be individually addressed. In
this sense, one needs to use collective operations, and spin
squeezing is one of the most successful approaches for controlling
such systems. For example, creating spin squeezing of an atomic
ensemble could result in precision measurements based on many-atom
spectroscopy~\cite{Wineland}. Therefore, we can also regard spin
squeezing as a quantum resource since for more than two particles
it behaves as two-particle entanglement in controlling and
detecting quantum correlations. On this quantum resource, we need
to further consider the effects of
decoherence~\cite{Deco,Ozdemir}. Thus, it is important to study
the environment-induced decoherence effects on both spin squeezing
and multipartite
entanglement~\cite{SimonKempe,Dur,Carvalho,JangNonlocality,Aolita,Lopez,Xiayj,Ficek,Guhne,Vedral,Werlang}.
A decaying time evolution of the spin squeezing under
decoherence~\cite{SimonKempe,Stockton,Laurat,Liyun} can be used to
analyze whether this quantum resource is robust.

In this article we address this problem by calculating three
spin-squeezing parameters  for a spin ensemble in a collective
excited state. We study the time evolution of spin squeezing under
local decoherence, acting independently and equally on each spin.
Here, the irreversible processes are modelled as three decoherence
channels: the amplitude damping, pure dephasing and depolarizing
channels. We find that, similar to the sudden death of pairwise
entanglement~\cite{Yuting}, spin squeezing can also suddenly
vanish with different lifetimes for some decoherence channels,
showing in general different vanishing times in multipartite
correlations in quantum many-body systems. Thus, similar to the
discovery of pairwise entanglement sudden death
(ESD)~\cite{Yuting}, the spin-squeezing sudden death (SSSD) occurs
due to decoherence. We will see that for some initial states, the
SSSD never occurs under the amplitude-damping channel. We also
give analytical expressions for the vanishing time of spin
squeezing and pairwise entanglement. The ESD has been tested
experimentally~\cite{Almeida,Laurat} and we also expect that the
SSSD can also be realized experimentally.

This article is organized as follows. In Sec.~II, we introduce the
initial state from the one-axis twisting Hamiltonian and then, in
Sec.~III, the decoherence channels. In Sec.~IV, we list three
parameters of spin squeezing and discuss the relations among them.
For a necessary comparison, the concurrence is also calculated. We
also study  initial-state squeezing. In Sec.~V, we study three
different types of spin squeezing and concurrence under three
different decoherence channels. Both analytical and numerical
results are given. We conclude in Sec.~VI.

\section{Initial state}

We consider an ensemble of $N$ spin-1/2 particles with ground
state $|1\rangle $ and excited state $|0\rangle.$ This system has
exchange symmetry, and its dynamical properties can be described
by the collective operators
\begin{equation}
J_{\alpha }=\sum_{k=1}^N
j_{k\alpha}=\frac{1}{2}\sum_{k=1}^{N}\sigma _{k\alpha }
\label{qqq}
\end{equation}
for $\alpha =x,y,z.$ Here, $\sigma _{k\alpha }$ are the Pauli
matrices for the $k$th qubit. To study the decoherence of spin
squeezing, we choose a state which is initially squeezed. One
typical class of such spin-squeezed states is the one-axis
twisting collective spin state~\cite{KU},
\begin{equation}
|\Psi (\theta_0 )\rangle _{0}=e^{-i\theta_0
J_{x}^{2}/2}|1\rangle^{\otimes N}=e^{-i\theta_0 J_{x}^{2}/2}|{\bf
1}\rangle , \label{initial}
\end{equation}
which could be prepared by the one-axis twisting Hamiltonian
\begin{equation}
H=\chi J_{x}^{2},
\end{equation}
where
\begin{equation}\label{angle}
\theta_0=2\chi t
\end{equation}
is the {\it one-axis twist angle} and $\chi$ is the coupling
constant. For this state, it was proved~\cite{WangSanders} that
the spin squeezing $\xi _{1}^{2}$~\cite{KU} and the concurrence
$C_{0}$~\cite{Conc} are equivalent since there exists a linear
relation $$\xi_1^2=1-(N-1)C_{0}$$ between them. Physically, they
occur and disappear simultaneously.  The spin squeezing of this
state can be generated and stored in,  e.g., a two-component
Bose-Einstein condensate~\cite{Jingr}.

\subsection{Initial-state symmetry}

The initial state has an obvious symmetry resulting from
Eq.~(\ref{initial}), the so-called even-parity symmetry, which
means that only even excitations of spins occur in the state.
Since $J_{\alpha }$ define an angular-momentum spinor
representation of SO(3), the general definitions of spin squeezing
for abstract operators $J_{x},J_{y},$ and $J_{z}$ can work well by
identifying $N/2$ with the highest weight $J$, which corresponds
to the collective ground state
\begin{equation}
|J,-J\rangle =|1\rangle^{\otimes N}\equiv |{\bf 1}\rangle
\end{equation}
indicating that all spins are in the ground state. The symmetric
space is generated by the collective operator $$J_{+}=
\frac12\sum_{k=1}^{N}\sigma _{k+} $$ acting on the collective
ground state. Here, $$\sigma_{k\pm}=\frac12(\sigma_{kx}\pm
i\sigma_{ky})$$. In others words, the state is in the maximally
symmetric space spanned by the Dicke states. So, the $N$ spin-1/2
system behaves like a larger spin-$N/2$ system. It can be proved
that any pure state with exchange symmetry belongs to the
above-mentioned symmetric space, but for mixed states the state
space can be extended to include a space beyond the symmetric
one~\cite{Molmer}. In the following discussions, we focus on such
an extended space.

In fact, after decoherence, not only the symmetric Dicke states
are populated, but also states with lower symmetry. So, it is not
sufficient to describe the system in only ($N+1)$-dimensional
space. Although the maximal symmetry is broken, the exchange
symmetry is not affected by the decoherence as each local
decoherence equally acts on each spin. In other words, a state
with exchange symmetry does not necessarily belong to the
maximally symmetric space.

With only the exchange symmetry, from Eq.~(\ref{qqq}),\ the global
expectations or correlations of collective operators are obtained as
\begin{eqnarray}
 \langle J_{\alpha }^{2}\rangle
&=&\frac{N}{4}+\frac{N(N-1)}{4}\langle \sigma
_{1\alpha }\sigma _{2\alpha }\rangle ,  \label{square4} \\
 \langle J_{-}^{2}\rangle &=&N(N-1)\langle \sigma _{1-}\sigma _{2-}\rangle ,
\label{s6} \\
 \langle \lbrack J_{x},J_{y}]_{+}\rangle &=&
\frac{N(N-1)}{4}\langle \lbrack \sigma _{1x},\sigma
_{2y}]_{+}\rangle.  \label{s5}
\end{eqnarray}
Furthermore, it follows from Eq.~(\ref{square4}) that
\begin{align}
& \langle J_{x}^{2}+J_{y}^{2}\rangle =\frac{N}{2}+\frac{N(N-1)}{2}\langle
\sigma _{1+}\sigma _{2-}+\sigma _{1-}\sigma _{2+}\rangle ,  \label{square2}
\\
& \langle J_{x}^{2}+J_{y}^{2}+J_{z}^{2}\rangle
=\frac{N^{2}}{4}\left[ \frac{3}{N}+\left( 1-\frac{1}{N}\right)
\langle \vec{\sigma}_{1}\cdot \vec{\sigma}_{2}\rangle \right] .
\label{square3}
\end{align}
These equations show the relations between the global and local
expectations and correlations, which are useful in the following
calculations.

\section{Decoherence channels and examples of their implementations}

Having introduced the initial state, now we discuss three typical
decoherence channels: the amplitude-damping channel (ADC), the
phase-damping channel (PDC), and the depolarizing channel (DPC).

These channels are prototype models of dissipation relevant in
various experimental systems. They provide ``a revealing
caricature of decoherence in realistic physical situations, with
all inessential mathematical details stripped
away''~\cite{Preskill}. But yet this ``caricature of decoherence''
leads to theoretical predictions being often in good agreement
with experimental data. Examples include multiphoton systems, ion
traps, atomic ensembles, or a solid-state spin systems such as
quantum dots or NV diamonds, where qubits are encoded in electron
or nuclear spins.

Here, we briefly describe only a few of such implementations.

\subsection{Amplitude-damping channel}

The ADC is defined as
\begin{equation}
{\cal E}_\text{ADC}(\rho)=E_0\rho E_0^\dagger + E_1\rho
E_1^\dagger, \label{ADC}
\end{equation}
where
\begin{equation}
E_{0}=\sqrt{s}\,|0\rangle \langle 0|+|1\rangle \langle 1|,\quad
E_{1}=\sqrt{p}\,|1\rangle \langle 0|  \label{kraus1}
\end{equation}
are the Kraus operators, $p=1-s$, $s=\exp (-\gamma t/2)$, and
$\gamma $ is the damping rate. In the Bloch representation, the
ADC squeezes the Bloch sphere into an ellipsoid and shifts it
toward the north pole. The radius in the $xy$ plane is reduced by
a factor $\sqrt{s}$, while in the $z$ direction it is reduced by a
factor $s$.

The ADC is a prototype model of a dissipative interaction between
a qubit and its environment. For example, the ADC model can be
applied to describe the spontaneous emission of a photon by a
two-level system into an environment of photon or phonon modes at
zero (or very low) temperature in (usually) the weak Born-Markov
approximation. The ADC can also describe processes contributing to
$T_1$ relaxation in spin resonance at zero temperature. Note that
by introducing an ``upward'' decay (i.e, a decay toward the south
pole of the Bloch sphere), in addition to the standard
``downward'' decay, the ADC can be used to describe dissipation
into the environment also at finite temperature.

The ADC acting on a system qubit in an unknown state $\rho$ can be
implemented in a two-qubit circuit performing a rotation
$R_y(\theta)$ of an ancilla qubit (initially in the ground state)
controlled by the system qubit and followed by a controlled-NOT
(CNOT) gate on the system qubit controlled by the ancilla
qubit~\cite{Nielsen}. The parameter $\theta$ is simply related to
the probability $p$ in Eq.~(\ref{ADC}). The ancilla qubit, which
models the environment, is measured after the gate operation.

The ADC-induced sudden vanishing of entanglement was first
experimentally demonstrated for polarization-encoded
qubits~\cite{Almeida}. For this reason let us shortly describe
this optical implementation of the ADC. It is based on a
Sagnac-type ring interferometer composed of a polarizing beam
splitter and a half-wave plate at an angle corresponding to the
parameter $p$ in Eq.~(\ref{ADC}). The beam splitter separates an
incident beam (being in a superposition of states with horizontal,
$|H\>$, and vertical, $|V\>$, polarizations) into spatially
distinct counter propagating light beams. The $H$ component leaves
the interferometer unchanged. But the $V$ component is rotated in
the wave plate, which corresponds to probabilistic damping into
the $H$ component. Then, at the exit from the interferometer, this
component is probabilistically transmitted or reflected from the
beam splitter. So it is cast into two orthogonal spatial modes
corresponding the reservoir states with and without excitation.

The action of the ADC can be represented by an interaction
Hamiltonian~\cite{Nielsen}: $H\sim  a b^\dagger + a^\dagger b$,
where $a$ ($a^\dagger$) and $b$ ($b^\dagger$) are annihilation
(creation) operators of the system and environment oscillators,
respectively. In more general models of damping, a single
oscillator $b$ of the reservoir is replaced by a finite or
infinite collection of oscillators $\{b_n\}$ coupled to the system
oscillator with different strengths (see, e.g.,
Ref.~\cite{Louisell,Leibfried03}). For the example of quantum
states of motion of ions trapped in a radio-frequency (Paul) trap,
the amplitude damping can be modeled by coupling an ion to the
motional amplitude reservoir described by the above
multioscillator Hamiltonian~\cite{Leibfried03}. The
high-temperature reservoir is possible to simulate by applying (on
trap electrodes) a random uniform electric field with spectral
amplitude at the ion motional
frequency~\cite{Myatt00,Turchette00}.  The zero-temperature
reservoir can be simulated by laser cooling combined with
spontaneous Raman scattering~\cite{Poyatos}.

\subsection{Phase-damping channel}

The PDC is a prototype model of dephasing or pure decoherence,
i.e., loss of coherence of a two-level state without any loss of
system's energy. The PDC is described by the map
\begin{equation}
{\cal E} _{\text{PDC}}(\rho )=s\rho +p\left( \rho _{00}|0\rangle
\langle 0|+\rho _{11}|1\rangle \langle 1|\right) \label{PDC},
\end{equation}
and obviously the three Kraus operators are given by
\begin{equation}
E_{0}=\sqrt{s}\,\openone,\; E_{1}=\sqrt{p}\,|0\rangle \langle
0|,\; E_{2}=\sqrt{p}\,|1\rangle \langle 1|,  \label{kraus2}
\end{equation}
where $\openone$ is the identity operator. For the PDC, there is
no energy change and a loss of decoherence occurs with probability
$p.$ As a result of the action of the PDC, the Bloch sphere is
compressed by a factor $(1-2p)$ in the $xy$ plane.

In analogy to the ADC, the PDC can be considered as an interaction
between two oscillators (modes) representing system and
environment as described by the interaction Hamiltonian: $H\sim
a^\dagger a(b^\dagger + b)$~\cite{Nielsen}. In more general
phase-damping models, a single environmental mode $b$ is usually
replaced by an infinite collection of modes $b_n$ coupled, with
various strengths, to mode $a$.

It is evident that the action of the PDC is nondissipative. It
means that, in the standard computational basis $|0\>$ and $|1\>$,
the diagonal elements of the density matrix $\rho$ remain
unchanged, while the off-diagonal elements are suppressed.
Moreover, the qubit states $|0\>$ and $|1\>$ are also unchanged
under the action of the PDC, although any superposition of them
(i.e., any point in the Bloch sphere, except the poles) becomes
entangled with the environment.

The PDC can be interpreted as elastic scattering between a
(two-level) system and a reservoir. It is also a model of coupling
a system with a noisy environment via a quantum nondemolition
(QND) interaction. Note that spin squeezing of atomic ensembles
can be generated via QND
measurements~\cite{Kuzmich99,Takahashi99,Kuzmich00,Julsgaard01,Schleier,Appel,Takano}.
So modeling the spin-squeezing decoherence via the PDC can be
relevant in this context.

The PDC is also a suitable model to describe $T_2$ relaxation in
spin resonance. This in contrast to modeling $T_1$ relaxation via
the ADC.

A circuit modeling the PDC can be realized as a simplified version
of the circuit for the ADC, discussed in the previous subsection,
obtained by removing the CNOT gate~\cite{Nielsen}. Then, the angle
$\theta$ in the controlled rotation gate $R_y(\theta)$ is related
to the probability $p$ in Eq.~(\ref{PDC}).

The sudden vanishing of entanglement under the PDC was first
experimentally observed in Ref.~\cite{Almeida}. This optical
implementation of the PDC was based on the same system as the
above-mentioned Sagnac interferometer for the ADC but with an
additional half-wave plate at a $\pi/4$ angle in one of the
outgoing modes.

Some specific kinds of PDCs can be realized in a more
straightforward manner. For example, in experiments with trapped
ions, the motional PDC can be implemented just by modulating the
trap frequency, which changes the phase of the harmonic motion of
ions~\cite{Myatt00,Turchette00} (for a review see
Ref.~\cite{Leibfried03} and references therein).

\subsection{Depolarizing channel}

The definition of the DPC is given via the map
\begin{align}
{\cal E }_{\text{DPC}}(\rho )& =\sum_{i=0}^{3}E_{k}\rho
E_{k}^{\dagger },
\\
& =(1-p^{\prime })\rho +\frac{p^{\prime }}{3}(\sigma _{x}\rho
\sigma _{x}+\sigma _{y}\rho \sigma _{y}+\sigma _{z}\rho \sigma
_{z}),  \notag\label{DPC},
\end{align}
where
\begin{eqnarray}
E_{0} &=&\sqrt{1-p^{\prime }}\openone,
\quad E_{1}=\sqrt{\frac{p^{\prime }}{3}}\sigma _{x},  \notag \\
E_{2} &=&\sqrt{\frac{p^{\prime }}{3}}\sigma _{y},\quad \quad
E_{3}=\sqrt{\frac{p^{\prime }}{3}}\sigma _{z}, \label{kraus3}
\end{eqnarray}
are the Kraus operators. By using the following identity
\begin{equation*}
\sigma _{x}\rho \,\sigma _{x}+\sigma _{y}\rho \,\sigma _{y}+\sigma
_{z}\rho \,\sigma_{z}+\rho =2\openone,
\end{equation*}%
we obtain
\begin{equation}
{\cal E}_{\text{DPC}}(\rho )=s\rho +p\frac{\openone}{2},
\end{equation}
where $p ={4p^{\prime }}/{3}$. We see that for the DPC, the spin
is unchanged with probability $s=1-p$ or it is depolarized to the
maximally mixed state $\openone/2\,\ $with probability $p.$ It is
seen that due to the action of the DPC, the radius of the Bloch
sphere is reduced by a factor $s$, but its shape remains
unchanged.

Formally, the action of the DPC on a qubit in an unknown state
$\rho$ can be implemented in a three-qubit circuit composed of two
CNOT gates with two auxiliary qubits initially in mixed states
\begin{eqnarray}
  \rho_1=\openone/2,\quad \rho_2=(1-p)|00\rangle\langle
00|+p|11\rangle\langle 11|, \label{N1}
\end{eqnarray}
which model the environment. Qubit $\rho_2$ controls the other
qubits via the CNOT gates~\cite{Nielsen}.

The DPC map can also be implemented by applying each of the Pauli
operators $[\openone,\sigma_x,\sigma_y,\sigma_z]$ at random with
the same probability. Using this approach, optical DPCs have been
realized experimentally both in free space~\cite{Ricci04} and in
fibers~\cite{Karpinski08}, where qubits are associated with
polarization states of single photons. In Ref.~\cite{Ricci04}, the
DPC was implemented by using a pair of equal electro-optical
Pockels cells. One of them was performing a $\sigma_x$ gate and
the other a $\sigma_y$ gate. The simultaneous action of both
$\sigma_x$ and $\sigma_y$ corresponds to a $\sigma_y$ gate. The
cells were driven (with a mutual delay of $\tau/2$) by a
continuous-wave periodic square-wave electric field with a
variable pulse duration $\tau$, so the total depolarizing process
lasted 2$\tau$ for each period.

Analogous procedures can be implemented in other systems,
including collective spin states of atomic ensembles. The coherent
manipulation of atomic spin states by applying off-resonantly
coherent pulses of light is a basic operation used in many
applications~\cite{Julsgaard04}. We must admit that the standard
methods enable rotations in the Bloch sphere of only classical
spin states (i.e., coherent spin states). Nevertheless,
recently~\cite{Takano} an experimental method has been developed
to rotate also spin-squeezed states.

It is worth noting that in experimental realizations of
decoherence channels (e.g, in ion-trap
systems~\cite{Hannemann09}), sufficient resources for complete
quantum tomography are provided even for imperfect preparation of
input states and the imperfect measurements of output states from
the channels.

\section{Spin-squeezing definitions and concurrence}

Now, we discuss several parameters of spin squeezing and give
several relations among them. To compare spin squeezing with
pairwise entanglement, we also give the definition of concurrence.
We notice that most previous investigations on ESD of concurrence
were only carried out for two-particle system rather than for
two-particle subsystem embedded in a larger system. For the
initial states, spin-squeezing parameters and concurrence are also
given below.

\subsection{Spin-squeezing parameters and their relations}
\subsubsection{Definitions of spin squeezing}

There are several spin-squeezing parameters, but we list only
three typical and related ones as
follows~\cite{KU,Wineland,Sorensen,Toth}:
\begin{eqnarray}
\xi _{1}^{2} &=&\frac{4(\Delta J_{\vec{n}_\perp })_{\min }^{2}}{N},~~  \label{x1} \\
\xi _{2}^{2} &=&\frac{N^{2}}{4\langle \vec{J}\rangle ^{2}}\xi _{1}^{2},~~
\label{x2} \\
\xi _{3}^{2} &=&\frac{\lambda _{\min }}{\langle \vec{J}^{2}\rangle
-\frac{N}{2}}.  \label{x3}
\end{eqnarray}
Here, the minimization in the first equation is over all
directions denoted by $\vec{n}_\perp,$ perpendicular to the mean
spin direction $\langle \vec{J}\rangle /\langle \vec{J}^{2}\rangle
$; $\lambda _{\min }$ is the minimum eigenvalue of the
matrix~\cite{Toth}
\begin{equation}
\Gamma =(N-1)\gamma +\mathbf{C},  \label{gamma}
\end{equation}
where
\begin{equation}
\gamma _{kl}={C}_{kl}-\langle J_{k}\rangle \langle J_{l}\rangle
\;\;\text{for}\;\; k,l\in \{x,y,z\}=\{1,2,3\}, \label{comatrix}
\end{equation}
is the covariance matrix and ${\bf C}=[C_{kl}]$ with
\begin{equation}
{C}_{kl}=\frac{1}{2}\langle J_{l}J_{k}+J_{k}J_{l}\rangle
\label{cmatrix}
\end{equation}
is the global correlation matrix. The parameters $\xi _{1}^{2},
\xi _{2}^{2},$ and $\xi _{3}^{2}$ were defined by Kitagawa and
Ueda \cite{KU}, Wineland {\em et al.}~\cite{Wineland}, and
T\'{o}th {\it et al.}~\cite{Toth}, respectively. If $\xi
_{2}^{2}<1$ $(\xi _{3}^{2}<1),$ spin squeezing occurs, and we can
safely say that the multipartite state is
entangled~\cite{Sorensen,Toth}. Although we cannot say that the
squeezed state via the parameter $\xi_1^2$ is entangled, it is
indeed closely related to quantum entanglement~\cite{WangSanders}.

\subsubsection{Squeezing parameters for states with parity}

We know from Sec.~II.A that the initial state has an even parity
and that the mean spin direction is along the $z$ direction.
During the transmission through all the three decoherence channels
discussed here, the mean spin direction does not change. For
states with a well-defined parity (even or odd), the
spin-squeezing parameter $\xi _{1}^{2}$ was found to be
\cite{WangSanders}
\begin{equation}
\xi _{1}^{2}=\frac{2}{N}\left( \langle J_{x}^{2}+J_{y}^{2}\rangle
-|\langle J_{-}^{2}\rangle |\right).  \label{xixi1}
\end{equation}
Then, the parameter $\xi _{2}^{2}$ given by Eq.~(\ref{x2}) becomes
\begin{equation}
\xi _{2}^{2}=\frac{N^{2}\xi _{1}^{2}}{4\langle J_{z}\rangle
^{2}}=\frac{N\left( \langle J_{x}^{2}+J_{y}^{2}\rangle -|\langle
J_{-}^{2}\rangle |\right) }{2\langle J_{z}\rangle ^{2}}.
\end{equation}
For the third squeezing parameter (see Appendix A for the
derivation), we have
\begin{equation}
\xi _{3}^{2}=\frac{\min \left\{ \xi _{1}^{2},\varsigma
^{2}\right\} }{{4}{N^{-2}}\langle \vec{J}^{2}\rangle
-{2}{N^{-1}}},  \label{xixixi}
\end{equation}
where
\begin{equation}
\varsigma ^{2}=\frac{4}{N^{2}}\left[ N(\Delta J_{z})^{2}+\langle
J_{z}\rangle ^{2}\right] .  \label{zzz}
\end{equation}
Note that the first parameter $\xi _{1}^{2}$ becomes a key
ingredient for the latter two squeezing parameters ($\xi_2^2$ and
$\xi_3^2$).

\subsubsection{Spin-squeezing parameters in terms of local expectations}

For later applications, we now express the squeezing parameters in
terms of local expectations and correlations, and also examine the
meaning of $\varsigma ^{2}$, which will be clear by substituting
Eqs.~(\ref{qqq}) and (\ref{square4}) into Eq.~(\ref{zzz}),
\begin{eqnarray}
\varsigma ^{2} &=&1+\mathcal{C}_{zz}  \notag \\
&=&1+(N-1)\left( \langle \sigma _{1z}\sigma _{2z}\rangle -\langle \sigma
_{1z}\rangle \langle \sigma _{2z}\rangle \right) .  \label{zzzz}
\end{eqnarray}
Thus, the parameter $\varsigma ^{2}$ is simply related to the
correlation $\mathcal{C}_{zz}$ along the $z$ direction. A negative
correlation ${\cal C}_{zz}<0$ is equivalent to $\varsigma ^{2}<1.$
It is already known that the spin-squeezing parameter $\xi
_{1}^{2}$ can be written as \cite{Kitagawa}
\begin{equation}\label{perp}
\xi _{1}^{2}=1+(N-1)\mathcal{C}_{\vec{n}_{\perp }\vec{n}_{\perp
}},
\end{equation}
where $\mathcal{C}_{\vec{n}_{\perp }\vec{n}_{\perp }}$ is the
correlation function in the direction perpendicular to the mean
spin direction. So, the spin squeezing $\xi_1^2<1$ is equivalent
to the negative pairwise correlations $\mathcal{C}_{\vec{n}_{\perp
}\vec{n}_{\perp }}<0$~\cite{Kitagawa}.

Thus, from the above analysis, spin squeezing and negative
correlations are closely connected to each other. The parameter
$\varsigma ^{2}<1$ indicates that spin squeezing occurs along the
$z$ direction, and $\xi_1^{2}<1$ implies spin squeezing along the
direction perpendicular to the mean spin direction. Furthermore,
from Eq.~(\ref{xixixi}), a competition between the transverse and
longitudinal correlations is evident.

By substituting Eqs.~(\ref{s6}) and (\ref{square2}) to
Eq.~(\ref{xixi1}), one can obtain the expression of $\xi _{1}^{2}$
in terms of local correlations $\langle \sigma _{1+}\sigma
_{2-}\rangle $ and $\langle \sigma _{1-}\sigma _{2-}\rangle $ as
follows:
\begin{eqnarray}
\xi _{1}^{2} &=&1+(N-1)\langle \sigma _{1+}\sigma _{2-}+\sigma _{1-}\sigma
_{2+}\rangle  \notag \\
&&-2(N-1)|\langle \sigma _{1-}\sigma _{2-}\rangle |  \notag \\
&=&1+2(N-1)\langle \sigma _{1+}\sigma _{2-}\rangle -|\langle
\sigma _{1-}\sigma _{2-}\rangle |).  \label{xixixi1}
\end{eqnarray}
The second equality in Eq.~(\ref{xixixi1}) results from the
exchange symmetry. From Eqs.~(\ref{qqq}), (\ref{square3}), and
(\ref{zzzz}), one finds
\begin{eqnarray}
\xi _{2}^{2} &=&\frac{\xi _{1}^{2}}{\langle \sigma _{1z}\rangle ^{2}},~
\label{xixixi2} \\
\xi _{3}^{2} &=&\frac{\min \left\{ \xi
_{1}^{2},1+\mathcal{C}_{zz}\right\} }{(1-N^{-1})\langle
\vec{\sigma}_{1}\cdot \vec{\sigma}_{2}\rangle +{N^{-1}}}.
\label{third}
\end{eqnarray}
Thus, we have reexpressed the squeezing parameters in terms of
local correlations and expectations.

\subsubsection{New spin-squeezing parameters}

In order to characterize spin squeezing more conveniently, we
define the following squeezing parameters:
\begin{equation}
\zeta _{k}^{2}=\max (0,1-\xi _{k}^{2}), \; k\in \{1,2,3\}.
\label{zeta}
\end{equation}
This definition is similar to the expression of the concurrence
given below. Spin squeezing appears when $\zeta _{k}^{2}>0$, and
there is no squeezing when $\zeta _{k}^{2}$ vanishes. Thus, the
definition of the first parameter $\zeta_{1}^{2}$ has a clear
meaning, namely, it is the \emph{strength} of the negative
correlations as seen from Eq.~(\ref{perp}). The larger is
$\zeta_1^2$, the larger is the strength of the negative
correlation, and the larger of is the squeezing. More explicitly,
for the initial state, we have $\xi _{1}^{2}=1-(N-1)C_{0}$
\cite{WangSanders}, so $\zeta _{1}^{2}$ is just the rescaled
concurrence $\zeta_1^2=C_{r}(0)=(N-1)C_{0}$~\cite{Vidal}.

Here, we give a few comments on the spin-squeezing parameter $\xi
_{2}^{2}$, which represents a competition between $\xi _{1}^{2}$
and $\langle \sigma _{1z}\rangle ^{2}$: the state is squeezed
according to the definition of $\ \xi _{2}^{2}$ if $\xi
_{1}^{2}<\langle \sigma _{1z}\rangle ^{2}$. We further note
that~\cite{CPL}
\begin{equation}
\langle \sigma _{1z}\rangle ^{2}=1-2E_{L},
\end{equation}
where $E_{L}$ is the linear entropy of one spin and it can be used
to quantify the entanglement of pure states~\cite{Horodecki}. So,
there is a competition between the strength of negative
correlations and the linear entropy $2E_L$ in the parameter $\xi
_{2}^{2},$ and $\zeta _{1}^{2}>2E_{L}$ implies the appearance of
squeezing.

\subsection{Concurrence for pairwise entanglement}

It has been found that the concurrence is closely related to spin
squeezing~\cite{WangSanders}. Here, we consider its behavior under
various decoherence channels. The concurrence quantifying the
entanglement of a pair of spin-1/2 can be calculated from the
reduced density matrix. It is defined as~\cite{Conc}
\begin{equation}
{C}=\max(0,\lambda _1-\lambda _2-\lambda _3-\lambda _4),
\label{Cdef}
\end{equation}
where the quantities $\lambda _i$ are the square roots of the
eigenvalues in descending order of the matrix product
\begin{equation}
\varrho _{12}=\rho _{12}(\sigma _{1y}\otimes \sigma _{2y})\rho
_{12}^{*}(\sigma _{1y}\otimes \sigma _{2y}).  \label{varrho}
\end{equation}
In (\ref{varrho}), $\rho _{12}^{*}$ denotes the complex conjugate
of $\rho _{12}$.

The two-spin reduced density matrix for a parity state with the
exchange symmetry can be written in a block-diagonal
form~\cite{WangMolmer}
\begin{equation}
\rho _{12}=\left(
\begin{array}{cc}
v_{+} & u^{\ast } \\
u & v_{-}%
\end{array}%
\right) \oplus \left(
\begin{array}{cc}
w & y \\
y & w%
\end{array}%
\right) ,  \label{re}
\end{equation}
in the basis \{$|00\rangle ,|11\rangle ,|01\rangle ,|10\rangle $\}, where
\begin{eqnarray}
v_{\pm } &=&\frac{1}{4}\left( 1\pm 2\langle \sigma _{1z}\rangle +\langle
\sigma _{1z}\sigma _{2z}\rangle \right) ,  \label{r1} \\
w &=&\frac{1}{4}\left( 1-\langle \sigma _{1z}\sigma _{2z}\rangle \right) ,
\label{r3} \\
u &=&\langle \sigma _{1+}\sigma _{2+}\rangle ,  \label{rrr} \\
y &=&\langle \sigma _{1+}\sigma _{2-}\rangle .  \label{rr}
\end{eqnarray}
The concurrence is then given by \cite{Wootters2}
\begin{equation}
C=\max \left\{ 0,2\left( |u|-w\right) ,2(y-\sqrt{v_{+}v_{-}})\right\} .
\label{conc}
\end{equation}
From the above expressions of the spin-squeezing parameters and
concurrence, we notice that if we know the expectation $\langle
\sigma _{1z}\rangle$, and the correlations $\langle \sigma
_{1+}\sigma _{2-}\rangle ,$ $\langle \sigma _{1-}\sigma
_{2-}\rangle$, and $\langle \sigma _{1z}\sigma _{2z}\rangle,$ all
the squeezing parameters and concurrence can be determined. Below,
we will give explicit analytical expressions for them subject to
three decoherence channels.

\subsection{Initial-state squeezing and concurrence}

We will now investigate initial spin squeezing and pairwise
entanglement by using our results for the spin-squeezing
parameters and concurrence obtained in the last subsections. We
find that the third squeezing parameter $\xi_3^2$ is equal to the
first one $\xi_1^2$. The squeezing parameter $\xi _{1}^{2}$ is
given by (see Appendix B):
\begin{align}
\xi _{1}^{2}(0)&=1-C_{r}(0)  \notag \\
& =1-(N-1)C_{0},  \notag \\
& =1-2(N-1)(|u_{0}|-y_{0}),  \label{ccc1}
\end{align}
where
\begin{align}
C_{0}=&\frac{1}{4}{\LARGE \{}[(1-\cos ^{N-2}\theta_0 )^{2}+16\sin
^{2}{(\theta_0/2)} \cos ^{2N-4}{(\theta_0/2)}]^{\frac{1}2}
\notag \\
& -1+\cos ^{N-2}\theta_0 {\Large \}}
\end{align}
is the concurrence \cite{WangSanders}.

The parameter $\xi _{2}^{2}(0)$ is easily obtained, as we know
both $\xi _{1}^{2}(0)$ and $\langle \sigma _{1z}\rangle_0^{2}$
(\ref{sigmaz})$.$ For this state, following from
Eq.~(\ref{square3}), $\langle \vec{\sigma}_{1}\cdot
\vec{\sigma}_{2}\rangle_0 =1,$ and thus the third parameter given
by Eq.~(\ref{third}) becomes
\begin{align}
\xi _{3}^{2}(0)&=\min [\xi _{1}^{2}(0),\varsigma ^{2}(0)]\notag\\
&=\min [\{ 1-C_{r}(0),1+\mathcal{C}_{zz}(0)] ,  \label{thirdd}
\end{align}
where the correlation function is
\begin{equation}
\mathcal{C}_{zz}(0)=\frac{1}{2}\left( 1+\cos ^{N-2}\theta_0
\right) -\cos ^{2N-2}{(\theta_0/2)}\geq 0.  \label{c5}
\end{equation}
The proof of the above inequality is given in Appendix C.

As the correlation function $\mathcal{C}_{zz}(0)$ and the
concurrence $C_{r}(0)$ are always $\ge 0$, Eq.~(\ref{thirdd})
reduces to
\begin{equation}
\xi _{3}^{2}(0)=\xi _{1}^{2}(0)=1-C_{r}(0).
\end{equation}
So, for the initial state, the spin-squeezing parameters $\xi
_{3}^{2}(0)$ and $\xi _{1}^{2}(0)$ are equal or equivalently, we
can write $\zeta _{1}^{2}(0)=\zeta _{3}^{2}(0)=C_{r}(0)$ according
to the definition of parameter $\zeta _{k}^{2}$ given by
Eq.~(\ref{zeta}). Below we made a summary of results of this
section in Table I.

\begin{table*}[tbp]
\caption{Spin-squeezing parameters $\xi_1^2$~\cite{KU},
$\xi_2^2$~\cite{Wineland}, $\xi_3^2$~\cite{Toth} and concurrence
$C$~\cite{Conc} for arbitrary states (first two columns), states
with parity (third column). The squeezing parameters are also
expressed in terms of local expectations (fourth column) and in
terms of the initial rescaled concurrence $C_r(0)$ for initial
states (last column). Also, $C_0$ is the initial concurrence, and
other parameters are defined in the text.}
\begin{center}
\begin{tabular}{c||c|c|c|c}
\hline\hline Squeezing parameters& Definitions & States with
parity & In terms of local expectations &Initial state
\\ \hline\hline
\parbox{2 cm} {$\xi_{1}^{2}$} &
\parbox{3.7 cm} {\vspace{2mm} $\displaystyle\frac{4(\Delta J_{\vec{n}_\perp })_{\min
}^{2}}{N}$ \vspace{2mm}} &
\parbox{4 cm}{$\displaystyle\frac{2}{N}\left( \langle J_{x}^{2}+J_{y}^{2}\rangle
-|\langle J_{-}^{2}\rangle |\right)$}&
\parbox{4 cm}{$1+2(N-1)(y -|u|)$}&
\parbox{1.5 cm}{$1-C_r(0)$}

\\ \hline
\parbox{2 cm} {$\xi_{2}^{2}$} &
\parbox{3.7cm} {\vspace{2mm} $\displaystyle\frac{N^{2}}{4\langle
\vec{J}\rangle ^{2}}\xi _{1}^{2}$ \vspace{2mm}} &
\parbox{4 cm} {$\displaystyle\frac{N^{2}\xi _{1}^{2}}{4\langle J_{z}\rangle ^{2}}$} &
\parbox{4 cm}{$\displaystyle\frac{\xi _{1}^{2}}{\langle \sigma _{1z}\rangle ^{2}}$}&
\parbox{1.5 cm}{$\displaystyle\frac{1-C_r(0)}{\langle\sigma_{1z}\rangle_0^2}$}
\\ \hline

\parbox{2 cm} { $\xi_{3}^{2}$} &
\parbox{3.9cm} {\vspace{2mm} \ $\displaystyle
\frac{\lambda _{\min }}{\langle \vec{J}^{2}\rangle
-\displaystyle\frac{N}{2}} $ \vspace{2mm} } &
\parbox{4.2 cm} { $\displaystyle\frac{\min
\left\{ \xi _{1}^{2},\varsigma ^{2}\right\} }{{4}{N^{-2}}\langle
\vec{J}^{2}\rangle -{2}{N^{-1}}}$}&
\parbox{4 cm}{$\displaystyle\frac{\min \left\{ \xi _{1}^{2},1+\mathcal{C}_{zz}\right\} }{(1-N^{-1})\langle \vec{\sigma}_{1}\cdot \vec{\sigma}_{2}\rangle +{N^{-1}}} $}&
\parbox{1.5 cm}{$1-C_r(0)$}
\\ \hline

\parbox{2.5 cm} {\vspace{0.3cm}Concurrence $C$\vspace{0.3cm}} &
\parbox{3.7 cm} {$\max(0,\lambda_1-\lambda_2-\lambda_3-\lambda_4)$} &
\parbox{4.0 cm} {$2\max (0,|u|-w,y-\sqrt{v_{+}v_{-}})$}&
\parbox{4.0 cm}{ $2\max (0,|u|-w,y-\sqrt{v_{+}v_{-}})$} &
\parbox{1.5 cm}{$C_0$}
\\ \hline
\end{tabular}
\end{center}
\end{table*}

\section{Spin squeezing under decoherence}

Now we begin to study spin squeezing under three different
decoherence channels. From the previous analysis, all the
spin-squeezing parameters and the concurrence are determined by
some correlation functions and expectations. So, if we know the
evolution of them under decoherence, the evolution of any
squeezing parameters and pairwise entanglement can be calculated.

\subsection{Heisenberg approach}

We now use the Heisenberg picture to calculate the correlation
functions and the relevant expectations. A decoherence channel
with Kraus operators $K_{\mu }$ is defined via the map
\begin{equation}
{\cal E}(\rho )=\sum_{\mu }K_{\mu }\rho K_{\mu
}^{\dagger }.
\end{equation}
Then, an expectation value of the operator $A$ can be calculated
as $\langle A\rangle =$Tr$\left[A{\cal E}(\rho )\right] .$
Alternatively, we can define the following map,
\begin{equation}
{\cal E}^{\dagger }(\rho )=\sum_{\mu }K_{\mu }^{\dagger }\rho
K_{\mu }.
\end{equation}
It is easy to check that%
\begin{equation}\label{four}
\langle A\rangle =\text{Tr}\left[ A{\cal E} (\rho) \right]
=\text{Tr}\left[{\cal E}^{\dagger }(A)\rho \right] .
\end{equation}
So, one can calculate the expectation value via the above equation
(\ref{four}). This is very similar to the standard Heisenberg
picture.

\subsection{Amplitude-damping channel}

\begin{figure}[tbp]
\includegraphics[width=9cm,clip]{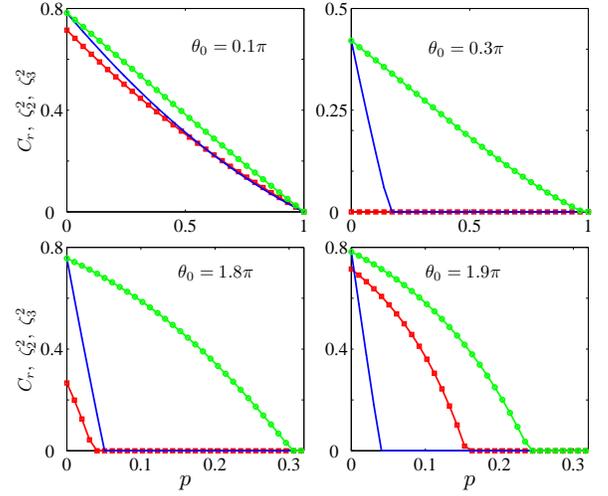}
\caption{(Color online) Spin-squeezing parameters
$\protect\zeta_2^2$ (red curve with squares), $\protect\zeta_3^2$
(top green curve with circles), and the concurrence $C_r$ (blue
solid curve) versus the decoherence strength $p=1-\exp(-\gamma t)$
for the amplitude-damping channel, where $\gamma$ is the damping
rate. Here, $\theta_0$ is the initial twist angle given by
Eq.~(\ref{angle}). In all figures, we consider an ensemble of
$N=12$ spins. Note that for a small initial twist angle $\theta_0$
(e.g., $\theta_0=0.1\pi$), the two squeezing parameters and the
concurrence all concur. For larger values of $\theta_0$, the
parameters $\zeta_2^2$, $\zeta_3^2$, and $C$ become quite
different and all vanish for sufficiently large values of the
decoherence strength.}
\end{figure}

\subsubsection{\protect\bigskip Squeezing parameters}
Based on the above approach and the Kraus operators for the ADC
given by Eq.~(\ref{kraus1}), we now find the evolutions of the
following expectations under decoherence (see Appendix D for
details)
\begin{subequations}
\begin{align}
\langle \sigma _{1z}\rangle =& \; s\langle \sigma _{1z}\rangle _{0}-p, \\
\langle \sigma _{1-}\sigma _{2-}\rangle =&\; s\langle \sigma
_{1-}\sigma
_{2-}\rangle_0 ,  \label{c2} \\
\langle \sigma _{1+}\sigma _{2-}\rangle =&\; s\langle \sigma
_{1+}\sigma
_{2-}\rangle_0 ,  \label{c3} \\
\langle \sigma _{1z}\sigma _{2z}\rangle =&\; s^{2}\langle \sigma
_{1z}\sigma _{2z}\rangle _{0}-2sp\langle \sigma _{1z}\rangle
_{0}+p^{2}.  \label{c44}
\end{align}
To determine the squeezing parameters and the concurrence, it is
convenient to know the correlation function $\mathcal{C}_{zz}$ and
the expectation $\langle \vec{\sigma}_{1}\cdot
\vec{\sigma}_{2}\rangle ,$ which can be determined from the above
expectations as follows:
\end{subequations}
\begin{align}
\langle \vec{\sigma}_{1}\cdot \vec{\sigma}_{2}\rangle =&1-s\, p\, x_{0}, \\
\mathcal{C}_{zz}=&s^{2}\left( \langle \sigma _{1z}\sigma _{2z}\rangle
_{0}-\langle \sigma _{1z}\rangle _{0}\langle \sigma _{2z}\rangle _{0}\right)
\notag  \label{c4} \\
=&s^{2}\mathcal{C}_{zz}(0),
\end{align}
where%
\begin{equation}
x_{0}=1+2\langle \sigma _{z}\rangle _{0}+\langle \sigma _{1z}\sigma
_{2z}\rangle _{0}.
\end{equation}
Substituting the relevant expectation values and the correlation
function into Eqs.~(\ref{xixixi1}), (\ref{xixixi2}), and
(\ref{third}) leads to the explicit expression of the
spin-squeezing parameters
\begin{eqnarray}
\xi _{1}^{2} &=&1-sC_{r}(0),  \label{xi1} \\
\xi _{2}^{2} &=&\frac{\xi _{1}^{2}}{\left( s\langle \sigma
_{1z}\rangle
_{0}-p\right) ^{2}},  \label{xix2} \\
\xi _{3}^{2} &=&\frac{\min \left\{\xi _{1}^{2},1+s^{2}\mathcal{C}%
_{zz}(0)\right\} }{1+({N}^{-1}-1)s\, p\, x_{0}}.  \label{xix3}
\end{eqnarray}
As the correlation function $\mathcal{C}_{zz}(0)\geq 0$, given by
Eq.~(\ref{c5})$,$ the third parameter can be simplified as
\begin{equation}
\xi _{3}^{2}=\frac{1-sC_{r}(0)}{1+({N}^{-1}-1)s\, p\, x_{0}}.
\end{equation}

Initially, the state is spin squeezed, i.e., $\xi _{1}^{2}(0)<1$
or $C_r(0)>0$. From Eq.~(\ref{xi1}), one can find that $\xi
_{1}^{2}<1$, except in the asymptotic limit of $p=1$. As we will
see below, for the PDC and DPC, $$\xi _{1}^{2}=1-s^{2}C_{r}(0).$$
Thus, we conclude that according to $\xi _{1}^{2}$, the initially
spin-squeezed state is always squeezed for $p\neq 1$, irrespective
of both the decoherence strength and decoherence models. In other
words, there exists no SSSD if we quantify spin squeezing by the
first parameter $\xi_1^2$.

\subsubsection{Concurrence}

In the expression (\ref{conc}) of the concurrence, there are three
terms inside the max function. The expression can be simplified to
(see Appendix E for details):
\begin{equation}
C_{r}=2(N-1)\max (0,|u|-w).  \label{sim}
\end{equation}
By using Eqs.~(\ref{r3}) and (\ref{c3}), one finds
\begin{align}
&2(|u|-w) \\
&\hspace{5mm}=2s|u_{0}|+\frac{s}{2}\left[ s-2+s\langle \sigma
_{1z}\sigma _{2z}\rangle
_{0}-2p\langle \sigma _{1z}\rangle _{0}\right] ) \notag\\
&\hspace{5mm}=sC_{0}-\frac{s\,p\,x_{0}}{2}.
\end{align}
So, we obtain the evolution of the rescaled concurrence as
\begin{equation}
C_{r}=\max \left[ 0,sC_{r}(0)-2^{-1}{(N-1)s\,p}\,x_{0}\right],
\label{cccc}
\end{equation}
which depends on the initial concurrence, expectation
$\langle\sigma_{1z}\rangle_0$, and correlation
$\langle\sigma_{1z}\sigma_{2z}\rangle_0$.

\subsubsection{Numerical results}

The numerical results for the squeezing parameters and concurrence
are shown in Fig.~1 for different initial values of  the twist
angle $\theta_0$, defined in Eq.~(\ref{angle}). For the smaller
value of $\theta_0$, e.g., $\theta_0=\pi/10$, we see that there is
no ESD and SSSD. All the spin squeezing and the pairwise
entanglement are completely robust against decoherence.
Intuitively, the larger is the squeezing, the larger is the
vanishing time. However, here, in contrast to this, no matter how
small are the squeezing parameters and concurrence, they vanish
only in the asymptotic limit. This results from the complex
correlations in the initial state and the special characteristics
of the ADC.

For larger values of $\theta_0$, as the decoherence strength $p$
increases, the spin squeezing decreases until it suddenly
vanishes, so the phenomenon of SSSD occurs. There exists a
critical value $p_{c},$ after which there is no spin squeezing.
The vanishing time of $\xi _{3}^{2}$ is always larger than those
of $\xi _{2}^{2}$ and the concurrence. We note that depending on
the initial state, the concurrence can vanish before or after $\xi
_{2}^{2}$. This means that in our model, the parameter $\xi
_{3}^{2}<1$ implies the existence of pairwise entanglement, while
$\xi _{2}^{2}$ does not.


\subsubsection{Decoherence strength $p_c$ corresponding to the SSSD}

From Eqs.~(\ref{xix2}), (\ref{xix3}), and (\ref{cccc}), the
critical value $p_{c}$ can be analytically obtained as
\begin{eqnarray}
p_{c}^{(k)} &=&\frac{x_kC_{r}(0)}{\left( N-1\right) x_{0}}, \quad (k=1,3) \label{eq1} \\
p_{c}^{(2)} &=&\frac{\langle \sigma_{1z}\rangle _{0}^{2}+C_{r}(0)-1}{%
1+2\langle \sigma_{1z}\rangle _{0}+\langle \sigma _{z}\rangle
_{0}^{2}},
\end{eqnarray}
where $x_1=2$ for the concurrence and $x_3=N$ for the squeezing
parameter $\zeta_{3}^{2}$.  The critical value $p_{c}^{(2)}$ is
for the second squeezing parameter $\zeta_{2}^{2}$. Here, $p_c$ is
related to the vanishing time $t_v$ via $p_c=1-\exp(-\gamma t_v)$.

In Fig.~2, we plot the critical values $p_c$ of the decoherence
strength versus $\theta_0$. The initial-state squeezing parameter
$\zeta_{1}^{2}$ is also plotted for comparison. For a range of
small values of $\theta_0$, the entanglement and squeezing are
robust to decoherence. The concurrence and parameter $\zeta
_{2}^{2}$ intersect. However, we do not see the intersections
between $\zeta _{3}^{2}$ and $\zeta _{2}^{2}$ or between
$\zeta_3^2$ and the concurrence. We also see that for the same
degree of squeezing, the vanishing times are quite different,
which implies that except for the spin-squeezing correlations,
other type of correlations exist. For large enough initial twist
angles $\pi\le \theta_0\le 2\pi$, the behavior of the squeezing
parameter $\xi_1^2$ is similar to those corresponding to
$p_{c}^{(1)}$ and $p_{c}^{(3)}$.

\begin{figure}[tbp]
\includegraphics[width=9cm,clip]{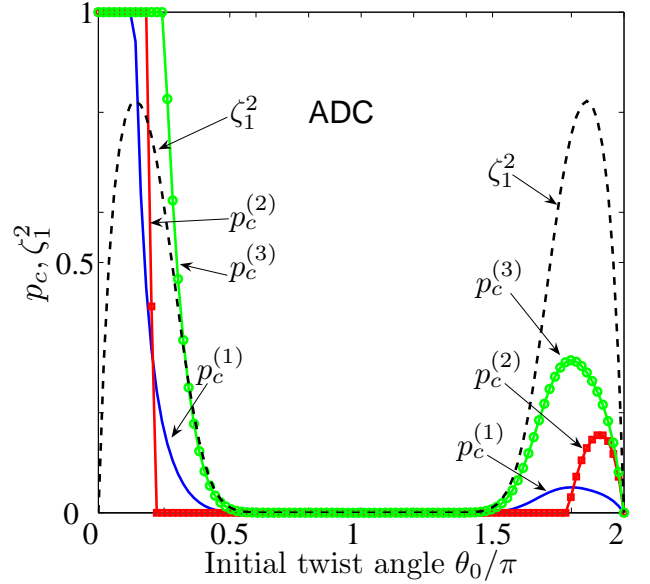}
\caption{(Color online) Critical values of the decoherence
strength $p_{c}^{(1)}$ (blue solid curve), $p_{c}^{(2)}$ (red
curve with squares), $p_{c}^{(3)}$ (green curve with circles), and
the squeezing parameter $\protect\zeta_1^2$ (black dashed curve)
versus the initial twist angle $\protect\theta_0$ given by
Eq.~(\ref{angle}) for the amplitude-damping channel, ADC. Here,
$p_c$ is related to the vanishing time $t_v$ via
$p_c=1-\exp(-\gamma t_v)$. At vanishing times, SSSD occurs. The
critical values $p_{c}^{(1)}$ , $p_{c}^{(2)}$, and $p_{c}^{(3)}$
correspond to the concurrence, squeezing parameter $\zeta_2^2$,
and $\zeta_3^2$, respectively.}
\end{figure}

\subsection{Phase-damping channel}

\subsubsection{Squeezing parameters and concurrence}
Now, we study the spin squeezing and pairwise entanglement under
the PDC. For this channel, the expectation values $\langle \sigma
_{z}^{\otimes n}\rangle $ are unchanged and the two correlations
$\langle \sigma _{1-}\sigma _{2-}\rangle $ and $\langle \sigma
_{1+}\sigma _{2-}\rangle $ evolve as (see Appendix D for
details)%
\begin{eqnarray}
\langle \sigma _{1-}\sigma _{2-}\rangle &=& s^{2}\langle \sigma
_{1-}\sigma _{2-}\rangle , \notag \\ \langle \sigma _{1+}\sigma
_{2-}\rangle &=& s^{2}\langle \sigma _{1+}\sigma _{2-}\rangle.
\label{evolve}
\end{eqnarray}
From the above equations and the fact $\langle \vec{\sigma}_{1}\cdot \vec{%
\sigma}_{2}\rangle _{0}=1$, one finds
\begin{align}
\langle \vec{\sigma}_{1}\cdot \vec{\sigma}_{2}\rangle
&=s^{2}\langle \sigma _{1x}\sigma _{2x}+\sigma _{1y}\sigma
_{2y}\rangle _{0}+\langle \sigma
_{1z}\sigma _{2z}\rangle _{0}  \notag \\
 &=s^{2}(1-\langle \sigma _{1z}\sigma _{2z}\rangle _{0})+\langle \sigma
_{1z}\sigma _{2z}\rangle _{0},  \label{sigmasigma} \\
\mathcal{C}_{zz}(p)&=\mathcal{C}_{zz}(0).  \label{sigmasigma2}
\end{align}
Therefore, from the above properties, we obtain the evolution of the
squeezing parameters,
\begin{eqnarray}
\xi _{1}^{2} &=&1-s^{2}C_{r}(0), \label{eee}\\
\xi _{2}^{2} &=&\frac{\xi _{1}^{2}}{\langle \sigma _{1z}\rangle _{0}^{2}}%
,~  \label{ee2}
\end{eqnarray}
and the third parameter becomes
\begin{eqnarray}
\xi _{3}^{2} &=&\frac{N\min \left[\xi _{1}^{2},1+\mathcal{C}%
_{zz}(0)\right] }{(N-{1})[ s^{2}+(1-s^{2})\langle \sigma_{1z}\sigma _{2z}\rangle_{0}] +1}  \label{ee3} \\
&=&\frac{N\xi _{1}^{2}}{(N-{1})[s^{2}+(1-s^{2})\langle \sigma
_{1z}\sigma _{2z}\rangle_{0}] +{1}}.
\end{eqnarray}
where we have used Eqs.~(\ref{sigmasigma}) and
(\ref{sigmasigma2}), and the property $\mathcal{C}_{zz}(0)\geq 0.$

From Eq.~(\ref{evolve}) and the simplified form of the concurrence
given by Eq.~(\ref{sim}), the concurrence is found to be
\begin{eqnarray}
C_{r} &=&\max \Big\{0,2(N-1)  \notag \\
&&\times \left[ s^{2}|u_{0}|-{4}^{-1}(1-\langle \sigma _{1z}\sigma
_{2z}\rangle _{0}\rangle )\right] \Big\}  \notag \\
&=&\max \left[0,s^{2}C_{r}(0)+\frac{a_{0}(s^{2}-1)}{2}\right].
\label{ee4}
\end{eqnarray}
where
\begin{align}
a_{0}=\left( N-1\right) (1-\langle \sigma _{1z}\sigma _{2z}\rangle
_{0}).
\end{align}
Thus, we obtained all time evolutions of the spin-squeezing
parameters and the concurrence. To study the phenomenon of SSSD,
we below examine the vanishing times.

\subsubsection{Decoherence strength $p_c$ corresponding to the SSSD}

The critical decoherence strengths $p_c$ can be obtained from Eqs.~(\ref{ee2}), (\ref{ee3}), and (%
\ref{ee4}) as follows:
\begin{eqnarray}
p_{c}^{(k)} &=&1-\left[ \frac{a_{0}}{x_kC_{r}(0)+a_{0}}\right]
^{\frac{1}{2}},
\label{eq2} \\
p_{c}^{(2)} &=&1-\left[ \frac{1-\langle \sigma _{1z}\rangle _{0}^{2}}{C_{r}(0)%
}\right]^{\frac{1}{2}},
\end{eqnarray}
where $k=1,3$ and $x_1=2, x_3=N$.
\begin{figure}[tbp]
\includegraphics[width=9cm,clip]{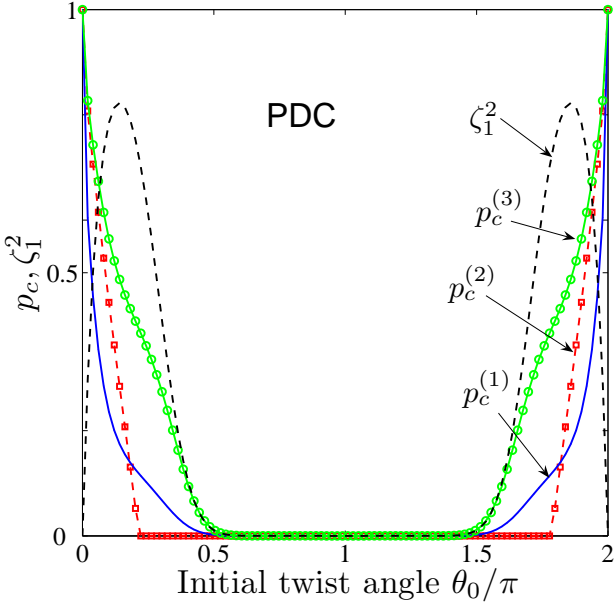}
\caption{(Color online) Same as in Fig.~2 but for the
phase-damping channel, PDC, instead of ADC. }
\end{figure}

In Fig. 3, we plot the decoherence strength $p_c$ versus the twist
angle $\theta_0$  of the initial state for the PDC. For this
decoherence channel, the critical value $p_c's$ first decrease
until they reach zero. Also, it is symmetric with respect to
$\theta_0 =\pi ,$ which is in contrast to the ADC. There are also
intersections between the concurrence and parameter $\xi
_{2}^{2},$ and the critical value $p_{c}^{(3)}$ is always larger
than $p_{c}^{(1)}$ and $p_{c}^{(2)}.$

\begin{table*}[tbp]
\caption{Analytical results for the time evolutions of all
relevant expectations, correlations, spin-squeezing parameters,
and concurrence, as well as the critical values $p_c$ of the
decoherence strength $p$. This is done for the three decoherence
channels considered in this work. For the concurrence $C$, we give
the expression for $C_r'$, which is related to the rescaled
concurrence $C_r$ via $C_r=\max(0,C_r')$.}
\begin{center}
\begin{tabular}{c||c|c|c}
\hline\hline &  Amplitude-damping channel & Phase-damping channel
& Depolarizing channel
\\
& (ADC) & (PDC) & (DPC)
\\ \hline\hline
\parbox{1.5 cm} {\vspace{0.3cm} $\langle\sigma_{1z}\rangle$\vspace{0.2cm} } &
\parbox{4 cm} {$s\langle\sigma_{1z}\rangle_0-p$} &
\parbox{4 cm}{$\langle\sigma_{1z}\rangle_0$}&
\parbox{4 cm}{$s\langle\sigma_{1z}\rangle_0$}
\\ \hline

\parbox{1.5 cm} {\vspace{0.3cm}$\langle\sigma_{1z}\sigma_{2z}\rangle$\vspace{0.3cm}} &
\parbox{4.3cm} {$s^2\langle\sigma_{1z}\sigma_{2z}\rangle_0-2sp\langle\sigma_{1z}\rangle_0+p^2$} &
\parbox{4 cm} {$\langle\sigma_{1z}\sigma_{2z}\rangle_0$}&
\parbox{4 cm}{$s^2\langle\sigma_{1z}\sigma_{2z}\rangle_0$}
\\ \hline

\parbox{1.5 cm} {\vspace{0.3cm}$\langle\sigma_{1+}\sigma_{2-}\rangle$\vspace{0.3cm}} &
\parbox{4cm} {$s\langle\sigma_{1+}\sigma_{2-}\rangle_0$} &
\parbox{4 cm} {$s^2\langle\sigma_{1+}\sigma_{2-}\rangle_0$}&
\parbox{4 cm}{$s^2\langle\sigma_{1+}\sigma_{2-}\rangle_0$}
\\ \hline

\parbox{1.5 cm} {\vspace{0.3cm}$\langle\sigma_{1-}\sigma_{2-}\rangle$\vspace{0.3cm}} &
\parbox{4cm} {$s\langle\sigma_{1-}\sigma_{2-}\rangle_0$} &
\parbox{4 cm} {$s^2\langle\sigma_{1-}\sigma_{2-}\rangle_0$}&
\parbox{4 cm}{$s^2\langle\sigma_{1-}\sigma_{2-}\rangle_0$}
\\ \hline

\parbox{1.5 cm} {\vspace{0.3cm}$\langle\vec{\sigma}_{1}\cdot\vec{\sigma}_{2}\rangle$\vspace{0.3cm}} &
\parbox{4cm} {$1-s\,p\,x_0$} &
\parbox{4 cm} {$s^2(1-\langle\sigma_{1z}\sigma_{2z}\rangle_0)+\langle\sigma_{1z}\sigma_{2z}\rangle_0$}&
\parbox{4 cm}{$s^2$}
\\ \hline

\parbox{1.5 cm} {\vspace{0.3cm}${\cal C}_{zz}$\vspace{0.3cm}} &
\parbox{4 cm} {$s^2{\cal C}_{zz}(0)$} &
\parbox{4 cm} {${\cal C}_{zz}(0)$}&
\parbox{4 cm}{$s^2{\cal C}_{zz}(0)$}
\\ \hline

\parbox{1.5 cm} {\vspace{0.3cm}$\xi_1^2$\vspace{0.3cm}} &
\parbox{4cm} {$1-sC_r(0)$} &
\parbox{4 cm} {$1-s^2C_r(0)$}&
\parbox{4 cm}{$1-s^2C_r(0)$}
\\ \hline

\parbox{1.5 cm} {\vspace{0.3cm}$\xi_2^2$\vspace{0.3cm}} &
\parbox{4cm} {$\displaystyle\frac{1-sC_r(0)}{(s\langle\sigma_{1z}\rangle_0-p)^2}$} &
\parbox{4 cm} {\vspace{0.15cm}$\displaystyle\frac{1-s^2C_r(0)}{\langle\sigma_{1z}\rangle_0^2}$\vspace{0.15cm}}&
\parbox{4 cm}{$\displaystyle\frac{1-s^2C_r(0)}{s^2\langle\sigma_{1z}\rangle_0^2}$}
\\ \hline

\parbox{1.5 cm} {\vspace{0.3cm}$\xi_3^2$\vspace{0.3cm}} &
\parbox{4cm} {$\displaystyle\frac{1-sC_r(0)}{1+(N^{-1}-1)s\,p\,x_0}$} &
\parbox{6 cm} {\vspace{0.15cm}$\displaystyle\frac{1-s^2C_r(0)}{(1-N^{-1})[s^2+(1-s^2)\langle\sigma_{1z}\sigma_{2z}\rangle_0]+N^{-1}}$\vspace{0.2cm}}&
\parbox{4 cm}{$\displaystyle\frac{1-s^2C_r(0)}{(1-N^{-1})s^2+N^{-1}}$}
\\ \hline

\parbox{1.5 cm} {\vspace{0.3cm}$C_r'$\vspace{0.3cm}} &
\parbox{4cm} {$sC_{r}(0)-(N-1)s\,p\,x_{0}/2$} &
\parbox{4 cm} {$s^{2}C_{r}(0)+{a_{0}(s^{2}-1)}/{2}$}&
\parbox{4.3 cm}{$s^{2}C_{r}(0)+(N-1)(s^{2}-1)/2$}
\\ \hline

\parbox{1.5 cm} {\vspace{0.3cm}$p_{c}^{(1)}$\vspace{0.3cm}} &
\parbox{4cm} {$\displaystyle\frac{2C_{r}(0)}{\left( N-1\right) x_{0}}$} &
\parbox{4 cm} {$\displaystyle 1-\left(
\frac{a_{0}}{2C_{r}(0)+a_{0}}\right)
^{\frac{1}{2}}$}&
\parbox{4 cm}{$\displaystyle 1-\left( \frac{N-1}{2 C_{r}(0)+N-1}\right)
^{\frac{1}2}$}
\\ \hline

\parbox{1.5 cm} {\vspace{0.3cm}$p_{c}^{(2)}$\vspace{0.3cm}} &
\parbox{4cm} {$\displaystyle\frac{\langle \sigma_{1z}\rangle _{0}^{2}+C_{r}(0)-1}{%
1+2\langle \sigma_{1z}\rangle _{0}+\langle \sigma _{z}\rangle
_{0}^{2}}$} &
\parbox{4 cm} {$\displaystyle 1-\left( \frac{1-\langle \sigma _{1z}\rangle _{0}^{2}}{C_{r}(0)%
}\right) ^{\frac{1}2}$}&
\parbox{4 cm}{$\displaystyle 1-\left( \frac{1}{C_{r}(0)+\langle \sigma _{1z}\rangle _{0}^{2}%
}\right) ^{\frac{1}2}$}
\\ \hline

\parbox{1.5 cm} {\vspace{0.3cm}$p_{c}^{(3)}\vspace{0.3cm}$} &
\parbox{4cm} {$\displaystyle\frac{NC_{r}(0)}{\left( N-1\right) x_{0}}$} &
\parbox{4 cm} {$\displaystyle 1-\left(
\frac{a_{0}}{NC_{r}(0)+a_{0}}\right)
^{\frac{1}{2}}$}&
\parbox{4 cm}{$\displaystyle 1-\left( \frac{N-1}{N C_{r}(0)+N-1}\right)
^{\frac12}$}
\\ \hline

\end{tabular}%
\end{center}
\end{table*}

\subsection{Depolarizing channel}

\begin{figure}[tbp]
\includegraphics[width=9cm,clip]{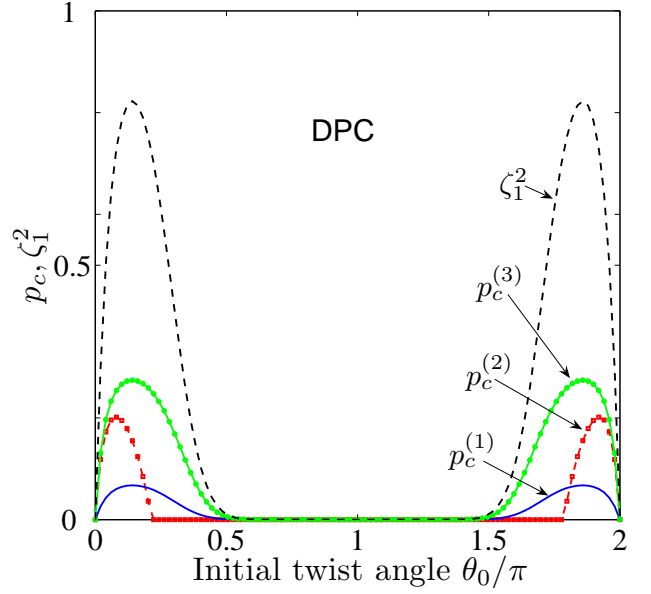}
\caption{(Color online) Same as in Fig.~2 but for the depolarizing
channel, DPC, instead of ADC.}
\end{figure}

\subsubsection{Squeezing parameters and concurrence}

The decoherence of the squeezing parameter defined by S\o rensen
{\it et al.}~\cite{Sorensen} has been studied in
Ref.~\cite{SimonKempe} for the DPC. It is intimately related to
the second squeezing parameter $\xi_2^2$. For the DPC, the
evolution of correlations $\langle \sigma _{1-}\sigma _{2-}\rangle
$ and $\langle \sigma _{1+}\sigma _{2-}\rangle $ are the same as
those of the DPC given by Eq.~(\ref{evolve}), and the expectations
$\langle \sigma _{1z}\rangle $ and $\langle \sigma _{1z}\sigma
_{2z}\rangle $ change as (see Appendix D).
\begin{eqnarray}
\langle \sigma _{1z}\rangle  &=&s\langle \sigma _{1z}\rangle _{0}, \\
\langle \sigma _{1z}\sigma _{2z}\rangle  &=&s^{2}\langle \sigma
_{1z}\sigma _{2z}\rangle_{0}.  \label{cccccc}
\end{eqnarray}
From these equations, we further have
\begin{align}
& \langle \vec{\sigma}_{1}\cdot \vec{\sigma}_{2}\rangle =s^{2}\langle \vec{%
\sigma}_{1}\cdot \vec{\sigma}_{2}\rangle _{0}=s^{2}, \\
& \mathcal{C}_{zz}=s^{2}\left( \langle \sigma _{1z}\sigma
_{2z}\rangle _{0}-\langle \sigma _{1z}\rangle _{0}\langle \sigma
_{2z}\rangle _{0}\right) =s^{2}\mathcal{C}_{zz}(0).
\end{align}
 The squeezing parameter $\xi_1^2$ is
given by Eq.~(\ref{eee}), and the other two squeezing parameters
are obtained as
\begin{eqnarray}
\xi _{2}^{2} &=&\frac{\xi _{1}^{2} }{s^{2}\langle \sigma
_{1z}\rangle
_{0}^{2}},~  \label{k1} \\
\xi _{3}^{2} &=&\frac{N\min \left\{\xi_{1}^{2} ,1+s^{2}\mathcal{C}%
_{zz}(0)\right\} }{(N-{1})s^{2}+{1}}  \notag \\
&=&\frac{N\xi _{1}^{2} }{(N-{1})s^{2}+{1}}. \label{k2}
\end{eqnarray}
By making use of Eqs.~( \ref{evolve}) and (\ref{cccccc}) and
starting from the simplified form of the concurrence (\ref{sim}),
we obtain
\begin{eqnarray}
C_{r} &=&\max
\left\{0,2(N-1)\left[s^{2}|u_{0}|-\textstyle{\frac14}(1-s^{2}\langle
\sigma
_{1z}\sigma _{2z}\rangle _{0})\right]\right\}   \notag \\
&=&\max \left[ 0,s^{2}C_{r}(0)+{2}^{-1}(N-1)(s^{2}-1)\right] .
\label{cb}
\end{eqnarray}
We observe that the concurrence is dependent only on the initial
value itself, not other ones.

\subsubsection{Decoherence strength $p_c$ corresponding to the SSSD}

From Eqs.~(\ref{cb}), (\ref{k1}), and (\ref{k2}), the vanishing
times are analytically calculated as
\begin{eqnarray}
p_{v}^{(k)} &=&1-\left[ \frac{N-1}{x_k C_{r}(0)+N-1}\right]
^{\frac 12}, \label{eq3}
\\
p_{v}^{(2)} &=&1-\left[ \frac{1}{C_{r}(0)+\langle \sigma _{1z}\rangle _{0}^{2}%
}\right]^{\frac 12},
\end{eqnarray}
where $k=1,3$ and $x_1=2, x_3=N$.

In Fig.~3, we plot the critical values $p_c$ versus the initial
twist angle $\theta_0$ for the DPC. For the DPC, the $p_c's$ first
increase until they reach their maxima and then decrease to zero.
Also, it is symmetric with respect to $\theta_0=\pi$, which is the
same as for the PDC. There are also intersections between the
concurrence and the parameter $\xi _{2}^{2}.$ Qualitatively, the
behaviors of $p_{c}^{(1)}$ and $p_{c}^{(3)}$ are the same as that
of the squeezing parameter $\zeta _{1}^{2}$. This implies that the
larger the squeezing, the larger is the critical value $p_c$.

The common features of these three decoherence channels are: (i)
The critical value $p_{v3}$ is always larger or equal than the
other two, namely, the spin-squeezing correlations according to
$\xi _{3}^{2}$ are more robust; (ii) there always exist two
intersections between the concurrence and the parameter $\xi
_{2}^{2},$ for $\theta_0$ from 0 to $2\pi $, irrespective of the
decoherence channels; (iii) when there is no squeezing (central
area of Figs. 2, 3, and 4), all vanishing times are zero. Table II
conveniently lists all the analytical results obtained in this
section.

\section{Conclusions and remarks}

To summarize, for a spin ensemble in a typical spin-squeezing
initial state under three different decoherence channels, we have
studied spin squeezing with three different parameters in
comparison with the pairwise entanglement quantified by the
concurrence. When the subsystems of the correlated system decay
asymptotically in time, the spin-squeezing parameter $\zeta
_{1}^{2}$ also decays asymptotically in time for all three types
of decoherence. However, for the other two squeezing parameters
$\zeta_2^2$ and $\zeta_3^2$, we find the appearance of
spin-squeezing sudden death and entanglement sudden death. The
global behaviors of the correlated state are markedly different
from the local ones. The spin-squeezing parameter $\zeta _{2}^{2}$
can vanish before, simultaneously, or after the concurrence, while
the squeezing parameter $\zeta _{3}^{2}$ is always the last to
vanish. This means that this parameter is more robust to
decoherence, and it can detect more entanglement than~$\xi_2^2$.

Our analytical approach for the vanishing times can be applied to
any initial quantum correlated states, not restricted to the
present one-axis twisted state. Moreover, for more complicated
channels, such as the amplitude-damping channel at finite
temperatures~\cite{Aolita} or the channel discussed in
Ref.~\cite{Lidar}, the method developed in this article can be
readily applied to study spin squeezing under these decoherence
channels.

Our investigations show the widespread occurrence of sudden death
phenomena in many-body quantum correlations. Since there exists
different vanishing times for different squeezing parameters, spin
squeezing offers a possible way to detect the total spin
correlation and their quantum fluctuations with distinguishable
time scales. The discovery of different lifetimes for various
spin-squeezing parameters means that, in some time region, there
still exists another quantum correlation when other quantum
correlations suddenly vanish. However, to determine which kind of
correlations will vanish, one possible approach is to further
invoke irreducible multiparty correlations~\cite{Zhou}, where the
multipartite correlations are classified in a series of
irreducible $k$ party ones. If we could obtain the time evolution
behaviors of such irreducible multipartite correlations in various
decoherence channels, we could classify lifetimes for the
spin-squeezing sudden death of various multipartite correlations
order by order.

\begin{acknowledgments}
We gratefully acknowledge partial support from the National
Security Agency, Laboratory of Physical Sciences, Army Research
Office, National Science Foundation under Grants Nos. 0726909, and
JSPS-RFBR 06-02-91200. X. Wang acknowledges support from the
National Natural Science Foundation of China under No. 10874151,
the National Fundamental Research Programs of China under Grant
No. 2006CB921205, and the Program for New Century Excellent
Talents in University (NCET). A.~M. acknowledges support from the
Polish Ministry of Science and Higher Education under Grant No. N
N202 261938.
\end{acknowledgments}

\begin{appendix}

\section{Spin-squeezing parameter $\protect\xi _{3}^{2}$ for states with
parity symmetry}

Here, we calculate the spin-squeezing parameter $\xi_3^2$ for
collective states with either even or odd parity symmetry. For
such states, we immediately have
\begin{equation}
\langle J_{x}\rangle =\langle J_{y}\rangle =\langle
J_{x}J_{z}\rangle =\langle J_{y}J_{z}\rangle =0
\end{equation}
as the operators change the parity of the state. Then, the mean
spin direction is along the $z$ direction and the correlation
matrix given by Eq.~(\ref{cmatrix}) is simplified to
\begin{equation}
\mathbf{C}=\left(
\begin{array}{ccc}
\langle J_{x}^{2}\rangle & C_{xy} & 0 \\
C_{xy} & \langle J_{y}^{2}\rangle & 0 \\
0 & 0 & \langle J_{z}^{2}\rangle%
\end{array}%
\right),
\end{equation}
where $C_{xy}=\langle \lbrack J_{x},J_{y}]_{+}\rangle/2$. From the
correlation matrix $\mathbf{C}$ and the definition of covariance
matrix $\gamma $ given by Eq.~(\ref{comatrix}), one finds
\begin{equation}
\Gamma =\left(
\begin{array}{ccc}
N\langle J_{x}^{2}\rangle & {N}C_{xy} & 0 \\
{N}C_{xy} & N\langle
J_{y}^{2}\rangle & 0 \\
0 & 0 & N(\Delta J_{z})^{2}+\langle J_{z}^{2}\rangle%
\end{array}%
\right).
\end{equation}
This matrix has a block-diagonal form and the eigenvalues of the
$2\times 2$ block are obtained as
\begin{equation}
\lambda _{\pm }=\frac{N}{2}\left( \langle J_{x}^{2}+J_{y}^{2}\rangle
\pm |\langle J_{-}^{2}\rangle |\right) .
\end{equation}
In deriving the above equation, we have used the relation%
\begin{equation}
J_{-}^{2}=J_{x}^{2}-J_{y}^{2}-i[J_{x},J_{y}]_{+}.
\end{equation}
Therefore, the smallest eigenvalue $\lambda _{\min }$ of $\Gamma $
is obtained as
\begin{equation}
\lambda _{\min }=\min \left(\lambda _{-},N(\Delta
J_{z})^{2}+\langle J_{z}^{2}\rangle \right),  \label{xixi2}
\end{equation}
where $\lambda _{-}$ differs from the squeezing parameter
$\xi_1^2$ given by Eq.~(\ref{xixi1}) by only a multiplicative
constant, as seen by comparing Eqs.~(\ref{xixi1}) and
(\ref{xixi2}). From Eqs.~(\ref{xixi2}) and (\ref{x3}), one finds
that the squeezing parameter $\xi_3^2$ is given by
Eq.~(\ref{xixixi}).

\section{Spin-squeezing parameters for the one-axis twisted state}

Here, we will use the Heisenberg picture to derive the relevant
expectations and spin-squeezing parameters for the initial
state~\cite{Molmer2,WangMolmer2}. To determine the spin-squeezing
parameter $\xi _{1}^{2}$ given by Eq.~(\ref{xixixi1}), one needs
to know the expectation $\langle \sigma_{1z}\rangle_0$, and
correlations $\langle \sigma _{1+}\sigma _{2-}\rangle_0$ and
$\langle \sigma _{1-}\sigma _{2-}\rangle_0$. We first consider the
expectation $\langle \sigma_{1z}\rangle_0$. For simplicity, we
omit the subscript $0$ in the following formulas.

\subsection{Expectation $\langle\sigma_{1z}\rangle$}

The evolution operator can be written as,
\begin{equation}
U=\exp({-i\chi tJ_{x}^{2}})=\exp\left({-i\theta
\sum_{k>l}j_{kx}j_{lx}}\right)
\end{equation}
up to a trivial phase, where $\theta=2\chi t$ given by
Eq.~(\ref{angle}). From this form, the evolution of $j_{1z}$ can
be obtained as
\begin{eqnarray}
U^{\dagger }j_{1z}U =j_{1z}\cos [ \theta j_x^{(2)}] +j_{1y}\sin [
\theta j_x^{(2)}],
\end{eqnarray}
where
\begin{equation}
j_{x}^{(k)}=\sum_{l=k}^N j_{lx}.
\end{equation}
Therefore, the expectations are
\begin{equation} \langle j_{1z}\rangle
=-{2}^{-1}\langle {\bf 1'}| \cos [ \theta j_x^{(2)}] |{\bf
1'}\rangle \label{jz1}
\end{equation}
since $\langle 1|j_{1y}|1\rangle =0.$ Here, $|{\bf 1'}\rangle
=|1\rangle _{2}\otimes ...\otimes |1\rangle _{N}.$ So, one can
find the following form for the expectation values
\begin{eqnarray}
\langle {\bf 1}|\cos \left[ \theta J_{x}\right] |{\bf 1}\rangle
&=&\left( \langle {\bf 1}|e^{i\theta J_{x}}|{\bf 1}\rangle
+ {\rm c.c.}\right) /2  \notag \\
&=&\left( \Pi _{k=1}^{N}\langle 1|e^{i\theta j_{kx}}|1\rangle + {\rm c.c.} \right) /2  \notag \\
&=&\cos ^{N}({\theta'}),  \label{jz2}
\end{eqnarray}
where $\theta'=\theta/2$ and $|{\bf 1}\rangle=|1\rangle^{\otimes
N}$.

By using Eqs.~(\ref{jz1}) and (\ref{jz2}), one gets
\begin{equation}\label{sigmaz}
\langle \sigma _{z}\rangle =-\cos ^{N-1}\left( {\theta' }\right).
\end{equation}

\subsection{\protect\bigskip Correlation $\langle \protect\sigma _{1+}%
\protect\sigma _{2-}\rangle $}

Since the operator $\sigma _{1x}\sigma _{2x}$ commutes with the
unitary operator $U,$ we easily obtain
\begin{equation}
\langle \sigma _{1x}\sigma _{2x}\rangle =0.  \label{xx}
\end{equation}
We now compute the correlations $\langle \sigma _{1z}\sigma
_{2z}\rangle .$ From the unitary operator,
\begin{eqnarray*}
&&\hspace*{-7mm} U^{\dagger }j_{1z}j_{2z}U\nonumber\\
&=&\left[j_{1z}\cos ( \theta
j_x^{(2)})+j_{1y}\sin ( \theta j_x^{(2)}) \right] \\
&&\times \left[j_{2z}\cos [ \theta (j_{1x}+j_{x}^{(3)})]
+j_{2y}\sin [ \theta (j_{1x}+j_{x}^{(3)})] \right] \\
&=&\left[j_{1z}\cos (\theta j_{2x})\cos(\theta
j_{x}^{(3)})-j_{1z}\sin
(\theta j_{2x})\sin (\theta j_{x}^{(3)})\right. \\
&&\left. +j_{1y}\sin (\theta j_{2x})\cos (\theta
j_{x}^{(3)})+j_{1y}\cos (\theta j_{2x})\sin (\theta j_{x}^{(3)})
\right] \\
&&\times \left[j_{2z}\cos (\theta j_{1x})\cos (\theta j_{x}^{(3)})
-j_{2z}\sin (\theta j_{1x})\sin (\theta j_{x}^{(3)})\right.
\\
&&\left. +j_{2y}\sin (\theta j_{1x})\cos (\theta j_{x}^{(3)})
+j_{2y}\cos (\theta j_{1x})\sin (\theta j_{x}^{(3)}) \right].
\end{eqnarray*}%
Although there are 16 terms after expanding the above equation,
only 4 terms survive when calculating $\langle s_{1z}s_{2z}\rangle
.$ We then have
\begin{eqnarray}\label{eef}
\langle j_{1z}j_{2z}\rangle
&=&\langle {\bf 1}|j_{1z}j_{2z}\cos ^{2}(\theta/2)\cos ^{2}(\theta j_{x}^{(3)})
\notag \\
&&-j_{1z}j_{2x}j_{2y}\sin (\theta )\sin ^{2}(\theta j_{x}^{(3)})
\notag \\
&&+4j_{1y}j_{1x}j_{2x}j_{2y}\sin ^{2}(\theta /2)\cos ^{2}(\theta j_{x}^{(3)})
\notag \\
&&-j_{1y}j_{1x}j_{2z}\sin (\theta )\sin ^{2}(\theta j_{x}^{(3)}) |{\bf 1}\rangle
\notag \\
&=&{4}^{-1}\langle {\bf 1}'|\cos ^{2}(\theta j_{x}^{(3)}) |{\bf 1}'\rangle
\notag \\
&=&{8}^{-1}\langle {\bf 1}'|\left[ 1+\cos (2\theta j_{x}^{(3)}
) \right]|{\bf 1}'\rangle
\notag \\
&=&{8}^{-1}\left[ 1+\cos ^{N-2}(\theta )\right],
\end{eqnarray}
where $|{\bf 1}'\rangle=|1\rangle_3\otimes...\otimes|1\rangle_N$.
The second equality in Eq.~(\ref{eef}) is due to the property
$j_{x}j_{y}=-j_{y}j_{x}={ij_z}/{2}$, and the last equality from
Eq.~(\ref{jz2}). Finally, from the above equation, one finds
\begin{equation}
\langle \sigma _{1z}\sigma _{2z}\rangle ={2}^{-1}\left( 1+\cos
^{N-2}\theta \right).  \label{zz}
\end{equation}
Due to the relation $\langle \sigma _{1x}\sigma _{2x}+\sigma
_{1y}\sigma _{2y}+\sigma _{1z}\sigma _{2z}\rangle =1$ for the
initial state, the correlation $\langle \sigma _{1y}\sigma
_{2y}\rangle $ is obtained from Eqs.~(\ref{xx}) and (\ref{zz}) as
\begin{equation}
\langle \sigma _{1y}\sigma _{2y}\rangle ={2}^{-1}\left( 1-\cos
^{N-2}\theta \right) .  \label{yy}
\end{equation}
Substituting Eqs.~(\ref{xx}) and (\ref{yy}) into the following
relations
\begin{equation*}
\sigma _{1x}\sigma _{2x}+\sigma _{1y}\sigma _{2y}=2\left( \sigma
_{1+}\sigma _{2-}+\sigma _{1-}\sigma _{2+}\right)
\end{equation*}
leads to one element of the two-spin reduced density matrix,
\begin{equation}\label{y0}
y_{0}=\langle \sigma _{1+}\sigma _{2-}\rangle ={8}^{-1}\left(
1-\cos ^{N-2}\theta \right) ,
\end{equation}
where the relation $\langle \sigma _{1+}\sigma _{2-}\rangle
=\langle \sigma _{1-}\sigma _{2+}\rangle $ is used due to the
exchange symmetry.

\subsection{Correlation $\langle \protect\sigma _{1-}\protect\sigma %
_{2-}\rangle $}

To calculate the correlation $\langle \sigma _{1-}\sigma
_{2-}\rangle ,$ due to the following relations%
\begin{eqnarray}
\sigma _{1x}\sigma _{2x}-\sigma _{1y}\sigma _{2y} &=&2\left( \sigma
_{1+}\sigma _{2+}+\sigma _{1-}\sigma _{2-}\right) ,  \label{sigma1} \\
i\left( \sigma _{1x}\sigma _{2y}+\sigma _{1y}\sigma _{2x}\right)
&=&2\left( \sigma _{1+}\sigma _{2+}-\sigma _{1-}\sigma
_{2-}\right) ,\quad  \label{sigma2}
\end{eqnarray}
we need to know the expectations $\langle j_{1x}j_{2y}\rangle .$ The
evolution of $j_{1x}j_{2y}$ is given by
\begin{eqnarray*}
U^{\dagger }s_{1x}s_{2y}U &=&j_{1x}\left\{j_{2y}\cos \left[ \theta
(j_{1x}+j_{x}^{(3)})\right] \right. \\
&&\left.\quad~~ -j_{2z}\sin \left[ \theta
(j_{1x}+j_{x}^{(3)})\right] \right\},
\end{eqnarray*}%
and the expectation is obtained as
\begin{eqnarray*}
\langle j_{1x}j_{2y}\rangle &=&{2}^{-1}\langle {\bf 1'}|j_{1x}\sin
\left[ \theta
(j_{1x}+j_{x}^{(3)})\right] |{\bf 1'}\rangle  \\
&=&{(4i)}^{-1}\langle {\bf 1'}|j_{1x}e^{i\theta j_{1x}}\Pi
_{k=3}^{N}e^{i\theta j_{kx}} \\
&&-j_{1x}e^{-i\theta j_{1x}}\Pi _{k=3}^{N}e^{-i\theta j_{kx}}|{\bf
1'}\rangle \\
&=&{(4i)}^{-1}{\cos ^{N-2}\left( {\theta'}{}\right) }\langle
1|j_{1x}e^{i\theta j_{1x}}-j_{1x}e^{-i\theta j_{1x}}|1\rangle  \\
&=&{2}^{-1}{\cos ^{N-2}\left( {\theta'}\right) }\langle 1|
j_{1x}\sin
(\theta j_{1x})|1\rangle  \\
&=&{4}^{-1}{\sin \left({\theta'}{}\right) \cos ^{N-2}\left(
\theta'\right) }
\end{eqnarray*}%
Here, $|{\bf 1'}\rangle=|1\rangle _{1}\otimes |1\rangle
_{3}\otimes ...\otimes |1\rangle _{N}$, where $|1\rangle_2$ is
absent. Moreover, $\langle j_{1y}j_{2x}\rangle =\langle
j_{1x}j_{2y}\rangle $ due to the exchange symmetry, and thus,
\begin{equation*}
\langle j_{1x}j_{2y}+j_{1y}j_{2x}\rangle =2^{-1}{\sin \left({\theta'}{}%
\right) \cos ^{N-2}\left({\theta'}{}\right) }.
\end{equation*}
For the initial state (\ref{initial}), we obtain the following
expectations \cite{KU,WangMolmer}
\begin{equation}\label{b12}
\langle \sigma _{1x}\sigma _{2y}+\sigma _{1y}\sigma _{2x}\rangle
=2\sin \left( {\theta'}{}\right) \cos ^{N-2}\left( {\theta
'}{}\right).
\end{equation}
The combination of Eqs.~(\ref{xx}), (\ref{yy}),  (\ref{sigma1}), (\ref{sigma2}%
), and (\ref{b12}) leads to the correlation
\begin{eqnarray}\label{u0}
u_{0} &=&\langle \sigma _{1-}\sigma _{2-}\rangle =-{8}^{-1}\left(
1-\cos
^{N-2}\theta \right)   \notag \\
&&-{i}{2}^{-1}\sin \left({\theta'}{}\right) \cos ^{N-2}\left({%
\theta' }{}\right).  \label{cr}
\end{eqnarray}
Substituting Eqs.~(\ref{y0}) and (\ref{u0}) to Eq.~(\ref{xixixi1})
leads to the expression of the squeezing parameter $\xi_1^2$ given
by Eq.~(\ref{ccc1}).

\section{Proof of ${\cal C}_{zz}(0)\ge 0 $}

To prove this, we will not use this specific function of the
initial twist angle $\theta$ as given by Eq.~(\ref{c5}), but use
the positivity of the reduced density matrix (\ref{re})$.$ We
first notice an identity
\begin{equation*}
\mathcal{C}_{zz}=4(v_{+}v_{-}-w^{2}),
\end{equation*}
which results from Eqs.~(\ref{r1}) and (\ref{r3}).  This is a key
step. Also there exists another identity
\begin{equation}
\label{e1} w_0=y_0
\end{equation}
as $\langle \vec{\sigma}_{1}\cdot \vec{\sigma}_{2}\rangle _{0}=1.$
From the positivity of the reduced density matrix (\ref{re}), one
has
\begin{equation*}
v_{0+}v_{0-}\geq |u_0|^{2}\geq y_0^{2}=w_0^{2},
\end{equation*}
where the second inequality follows from Eq.~(\ref{r3}) and the
last equality results from Eq.~(\ref{e1}). This completes the
proof.

\section{Derivation of the evolution of the correlations and expectations under decoherence}

For an arbitrary matrix
\begin{equation*}
A=\left(
\begin{array}{cc}
a & b \\
c & d%
\end{array}%
\right) ,
\end{equation*}
from the Kraus operators (\ref{kraus1}) for the ADC, it is
straightforward to find
\begin{eqnarray*}
{\cal E} (A) &=&\left(
\begin{array}{cc}
sa & \sqrt{s}b \\
\sqrt{s}c & d+pa%
\end{array}%
\right) , \\
{\cal E}^{\dagger }(A) &=&\left(
\begin{array}{cc}
sa+pd & \sqrt{s}b \\
\sqrt{s}c & d%
\end{array}%
\right) .
\end{eqnarray*}%
The above equations imply that
\begin{eqnarray*}
{\cal E} ^{\dagger }(\sigma _{\mu}) &=&\sqrt{s}\sigma _{\mu} \; \text{for}\; \mu=x,y, \\
{\cal E} ^{\dagger }(\sigma _{z}) &=&s \sigma _{z}-p.
\end{eqnarray*}%
As we considered independent and identical decoherence channels
acting separately on each spin, the evolution correlations and expectations in Eqs.~(%
\ref{c2}), (\ref{c3}), and (\ref{c44}) are obtained directly from
the above equations.

From the Kraus operators (\ref{kraus2}), the evolution of the
matrix $A$ under the PDC is obtained as
\begin{equation*}
{\cal E} (A)={\cal E} ^{\dagger }(A)=\left(
\begin{array}{cc}
a & sb \\
sc & d
\end{array}
\right) ,
\end{equation*}
from which one finds
\begin{eqnarray*}
{\cal E} ^{\dagger }(\sigma _{\mu}) &=&s \sigma _{\mu} \quad \text{for} \; \mu=x,y \\
{\cal E} ^{\dagger }(\sigma _{z}) &=&\sigma _{z}.
\end{eqnarray*}%
So expectations $\langle \sigma _{z}^{\otimes n}\rangle $ are
unchanged and Eq.~(\ref{evolve}) is obtained.

From the Kraus operators (\ref{kraus3}) of the DPC, the evolution
of the matrix $A$ is given by
\begin{eqnarray*}
{\cal E} (A) &=&{\cal E} ^{\dagger }(A) \\
&=&\left(
\begin{array}{cc}
as +\frac{p}{2}(a+d) & sb \\
sc & ds +\frac{p}{2}(a+d)%
\end{array}%
\right)
\end{eqnarray*}%
from which one finds%
\begin{eqnarray*}
{\cal E} ^{\dagger }(\sigma _{\alpha}) =s \sigma _{\alpha}\quad
\text{for}\; \alpha\in\{x,y,z\}.
\end{eqnarray*}%
Then, Eq.~(\ref{cccccc}) is obtained.

\section{Simplified form of the concurrence}

For our three kinds of decoherence channels, the concurrence
(\ref{conc}) can be simplified and given by
\begin{eqnarray}
C &=&\max \left\{ 0,2\left( |u|-w\right)
,2(y-\sqrt{v_{+}v_{-}})\right\}
\notag\\
&=&\max \left\{ 0,2\left( |u|-w\right) \right\} . \label{E1}
\end{eqnarray}
If one can prove
\begin{eqnarray}
|u|-y &\geq &0, \\
w-\sqrt{v_{+}v_{-}} &\leq &0,
\end{eqnarray}
then we obtain the simplified form shown in Eq.~(\ref{E1}). The
last inequality can be replaced by
\begin{equation}\label{c22}
w^{2}-v_{+}v_{-}\leq 0
\end{equation}
as $w$ and $v_{+}v_{-}$ are real.

We first consider the ADC channel. From Eqs.~(\ref{c2}), (\ref{c3}), and (\ref{c4}%
), one obtains
\begin{eqnarray}
|u|-y &=&s(|u_{0}|-y_{0})\geq 0, \\
w^{2}-v_{+}v_{-} &=&-\frac{1}{4}\mathcal{C}_{zz}=-\frac{s^{2}}{4}\mathcal{C}%
_{zz}(0)\leq 0.
\end{eqnarray}
where the inequalities result from Eqs.~(\ref{ccc1}) and
(\ref{c5}), respectively. So, the inequality (\ref{c22}) follows.

For the PDC, from Eq.~(\ref{evolve}) and fact that
$\langle\sigma_z^{\otimes n}\rangle$ is unchanged under
decoherence, the concurrence can also be simplified due to the
following properties:
\begin{eqnarray*}
|u|-y &=&s^{2}(|u_{0}|-y_{0})\geq 0, \\
w^{2}-v_{+}v_{-} &=&-\frac{1}{4}\mathcal{C}_{zz}(0)\leq 0.
\end{eqnarray*}
For the DPC, from Eqs.~(\ref{evolve}) and (\ref{cccccc}), one has
\begin{eqnarray}
|u|-y &=&s^{2}(|u_{0}|-y_{0})\geq 0, \\
w^{2}-v_{+}v_{-} &=&-\frac{s^{2}}{4}\mathcal{C}_{zz}(0)\leq 0.
\end{eqnarray}
So, again, the concurrence can be simplified to the form shown in
Eq.~(\ref{E1}). This completes the proof.

\end{appendix}


\begin{thebibliography}{99}
\bibitem{SSS}
A. Einstein, B. Podolsky, and R. Rosen, Phys. Rev. \textbf{47},
777 (1935); E. Schr\"{o}dinger, Naturwissenschaften \textbf{23},
807 (1935).

\bibitem{Nielsen} M. A. Nielsen and I. L. Chuang, \textit{Quantum
Computation and Quantum Information} (Cambridge University Press,
Cambridge, UK, 2000).

\bibitem{Shiyh} Y. H. Shih and C. O. Alley, \prl \textbf{61},
2921 (1988).

\bibitem{Polzik}
K. Hammerer, A.S. S\o rensen, E.S. Polzik,
eprint arXiv:0807.3358.

\bibitem{Lukin} A. Andr\'{e} and M. D. Lukin, \pra \textbf{65},
053819 (2002); M. D. Lukin, S. F. Yelin, and M. Fleischhauer, \prl
\textbf{84}, 4232 (2000).

\bibitem{Leibfried} D. Leibfried, M. D. Barrett,
T. Schaetz, J. Britton, J. Chiaverini, W. M. Itano, J. D. Jost, C.
Langer, and D. J. Wineland, Science \textbf{304}, 1476 (2004).

\bibitem{Panjw} J. W. Pan, M. Daniell, S. Gasparoni, G. Weihs, and A.
Zeilinger, \prl {\bf 86}, 4435 (2001).

\bibitem{Guogc} Y. F. Huang, X. L. Niu, Y. X. Gong, J. Li,
L. Peng, C. J. Zhang, Y. S. Zhang, and G. C. Guo, \pra {\bf 79},
052338 (2009).

\bibitem{Dujf} Z. Y. Xu, Y. M. Hu, W. L. Yang, M. Feng, and J. F. Du, \pra {\bf 80},
022335 (2009).

\bibitem{Appel}
J. Appel, P. J. Windpassinger, D. Oblak, U. B. Hoff, N. Kj\ae
rgaard, and E. S. Polzik,
Proc. Natl. Acad. Sci. USA {\bf 106}, 10960 (2009).

\bibitem{Andre}A. Andr\'{e}, A. S. S\o rensen, and M. D. Lukin, \prl {\bf 92},
230801 (2004).

\bibitem{Conc} W. K. Wootters, \prl {\bf 80}, 2245 (1998).

\bibitem{Neg} A. Peres, \prl \textbf{77} 1413 (1996); M.
Horodecki, P. Horodecki and R. Horodecki, Phys. Lett. A
\textbf{223}, 1 (1996); G. Vidal and R. F. Werner, \pra
\textbf{65} 032314 (2002).

\bibitem{Horodecki}R. Horodecki, P. Horodecki, M. Horodecki, and K. Horodecki,
\rmp {\bf 81}, 865 (2009).

\bibitem{Witness} A. Ac\'{\i}n, D. Bruss, M. Lewenstein, and A. Sanpera, %
\prl {\bf 87}, 040401 (2001); O. G\"{u}hne and G. T\'{o}th, Phys. Rep.
\textbf{474}, 1 (2009).

\bibitem{KU} M. Kitagawa and M. Ueda, \pra {\bf 47}, 5138 (1993).

\bibitem{Wineland} D. J. Wineland, J. J. Bollinger, W. M. Itano, and D. J.
Heinzen, \pra {\bf 50}, 67 (1994).

\bibitem{Sorensen} A. S\o rensen, L.-M. Duan, J. I. Cirac, and P. Zoller,
Nature (London) \textbf{409}, 63 (2001).

\bibitem{Toth}G. Toth, C. Knapp, O. G\"{u}hne, and H. J. Briegel, \prl {\bf 99}, 250405 (2007);
\pra \textbf{79}, 042334 (2009).

\bibitem{Korbicz} J. K. Korbicz, J. I. Cirac, and M. Lewenstein,
\prl {\bf
95}, 120502 (2005).

\bibitem{WangSanders} X. Wang and B. C. Sanders, \pra {\bf 68}, 012101
(2003).

\bibitem{Berman} C. Genes, P. R. Berman, and A. G. Rojo,
\pra \textbf{68}, 043809 (2003).

\bibitem{Fernholz}
T. Fernholz, H. Krauter, K. Jensen, J. F. Sherson, A. S. S\o
rensen, and E. S. Polzik,
\prl {\bf 101}, 073601 (2008).

\bibitem{Takano} T. Takano, M. Fuyama, R. Namiki, and Y. Takahashi,
\prl \textbf{102}, 033601 (2009); T. Takano, S. I. R. Tanaka, R.
Namiki, and Y. Takahashi,
eprint arXiv:0909.2423v1.

\bibitem{Deco} W. H. Zurek, Rev. Mod. Phys. \textbf{75}, 715 (2003).

\bibitem{Ozdemir}
S. K. \"Ozdemir, K. Bartkiewicz, Y. X. Liu, and A. Miranowicz,
\pra  {\bf 76}, 042325 (2007).

\bibitem{SimonKempe} C. Simon and J. Kempe, \pra \textbf{65}, 052327
(2002).

\bibitem{Dur} W. D\"{u}r and H. -J. Briegel, \prl {\bf 92}, 180403 (2004).

\bibitem{Carvalho} A. R. R. Carvalho, F. Mintert, and A. Buchleitner, \prl
\textbf{93}, 230501 (2004).

\bibitem{JangNonlocality} S. S. Jang, Y. W. Cheong, J. Kim, and H. W. Lee, %
\pra {\bf 74}, 062112 (2006).

\bibitem{Aolita} L. Aolita, R. Chaves, D. Cavalcanti, A. Ac\'{i}n, and L. Davidovich, \prl {\bf 100}, 080501 (2008);
L. Aolita, D. Cavalcanti, A. Ac\'{i}n, A. Salles, M. Tiersch, A.
Buchleitner, and F. de Melo, \pra {\bf 79}, 032322 (2009).

\bibitem{Lopez} C. E. L\'{o}pez, G. Romero, F. Lastra, E. Solano, and J. C. Retamal,
\prl {\bf 101}, 080503 (2008).

\bibitem{Xiayj} Z. X. Man, Y. J. Xia, and N. B. An, \pra {\bf 78},
064301 (2008); N. B. An and J. Kim, \pra {\bf 79}, 022303 (2009).

\bibitem{Ficek}
Z. Ficek and R. Tana\'{s}, \pra {\bf 77}, 054301
(2008).

\bibitem{Guhne} O. G\"{u}hne, F. Bodoky, and M. Blaauboer, \pra
\textbf{78}, 060301(R) (2008).

\bibitem{Vedral}
J. Maziero, L. C. C\'{e}leri, R. M. Serra, and V. Vedral,
\pra 80, 044102 (2009). 

\bibitem{Werlang} T. Werlang, S. Souza, F. F. Fanchini, and C. J. Villas
Boas, \pra \textbf{80}, 024103 (2009).

\bibitem{Stockton} J. K. Stockton, J. M. Geremia, A. C. Doherty, and H.
Mabuchi, \pra {\bf 67}, 022112 (2003).

\bibitem{Laurat}
J. Laurat, K. S. Choi, H. Deng, C. W. Chou, and H. J. Kimble,
\prl \textbf{99}, 180504 (2007).

\bibitem{Liyun} Y. Li, Y. Castin, and A. Sinatra, \prl {\bf 100}, 210401
(2008).

\bibitem{Yuting} T. Yu and J. H. Eberly, \prl {\bf 93}, 140404 (2004); Science \textbf{323}, 598 (2009).

\bibitem{Almeida}
M. P. Almeida, F. de Melo, M. Hor-Meyll, A. Salles, S. P. Walborn,
P. H. Souto Ribeiro, and L. Davidovich,
Science \textbf{316}, 579 (2007); A. Salles, F. de Melo, M. P.
Almeida, M. Hor-Meyll, S. P. Walborn, P. H. Souto Ribeiro, and L.
Davidovich,
\pra {\bf 78}, 022322 (2008).

\bibitem{Jingr} G. R. Jin and S. W. Kim, \prl {\bf 99}, 170405 (2007).

\bibitem{Molmer} J. Wesenberg and K. M\o lmer, \pra \textbf{65},
062304 (2002).

\bibitem{Preskill} J. Preskill, {\em Lecture Notes for Physics 219:
Quantum Information and Computation} (Caltech, Pasadena, CA,
1999), www.theory.caltech.edu/people/preskill/ ph229.

\bibitem{Louisell}
W. Louisell, {\em Quantum Statistical Properties of Radiation}
(Wiley, New York, 1974).

\bibitem{Leibfried03}
D. Leibfried, R. Blatt, C. Monroe, and D. Wineland,
Rev. Mod. Phys. {\bf 75}, 281 (2003).

\bibitem{Myatt00}
C. J. Myatt, B. E. King, Q. A. Turchette, C. A. Sackett, D.
Kielpinski, W. M. Itano, C. Monroe, and D. J. Wineland,  Nature
(London) {\bf 403}, 269 (2000).

\bibitem{Turchette00}
Q. A. Turchette, C. J. Myatt,  B. E. King,  C. A. Sackett, D.
Kielpinski, W. M. Itano, C. Monroe,  and D. J. Wineland,
Phys. Rev. A {\bf 62}, 053807 (2000).

\bibitem{Poyatos}
J. F. Poyatos, J. I. Cirac, and P. Zoller,
\prl {\bf 77}, 4728 (1996).

\bibitem{Kuzmich99}
A. Kuzmich, L. Mandel, J. Janis, Y. E. Young, R. Ejnisman, and N.
P. Bigelow,
\pra {\bf 60}, 2346 (1999).

\bibitem{Takahashi99}
Y. Takahashi, K. Honda, N. Tanaka, K. Toyoda, K. Ishikawa, and T.
Yabuzaki,
\pra {\bf 60}, 4974 (1999).

\bibitem{Kuzmich00}
A. Kuzmich, L. Mandel, and N. P. Bigelow,
\prl  {\bf 85}, 1594 (2000).

\bibitem{Julsgaard01}
B.  Julsgaard,  A.  Kozhekin,  and  E.  S.  Polzik,
Nature (London) {\bf 413}, 400 (2001).

\bibitem{Schleier}
M.  H.  Schleier-Smith,  I.  D.  Leroux,  and  V.  Vuleti\'c,
e-print quant-ph/0810.2582.

\bibitem{Ricci04}
M. Ricci, F. De Martini,  N. J. Cerf,  R. Filip,  J. Fiura\v{s}ek,
and C. Macchiavello,
\prl  {\bf 93}, 170501 (2004).

\bibitem{Karpinski08}
M. Karpi\'nski, C. Radzewicz, and K. Banaszek,
J. Opt. Soc. Am. B {\bf 25}, 668 (2008).

\bibitem{Julsgaard04}
B. Julsgaard, J. Sherson, J. I. Cirac, J. Fiura\v{s}ek, and E. S.
Polzik,
Nature (London) {\bf 432}, 482 (2004).

\bibitem{Hannemann09}
T. Hannemann, Ch. Wunderlich, M. Plesch, M. Ziman, and V.
Bu\v{z}ek,
e-print arXiv:0904.0923.

\bibitem{Kitagawa} D. Ulam-Orgikh and M. Kitagawa, \pra \textbf{64}, 052106 (2001).

\bibitem{Vidal}J. Vidal, G. Palacios, and R. Mosseri, \pra {\bf 69} 022107 (2004).

\bibitem{CPL}D. Yan, X. Wang, and L. A. Wu, Chin. Phys. Lett. {\bf 22}, 271 (2005).

\bibitem{WangMolmer} X. Wang and K. M\o lmer, Euro. Phys. J. D {\bf 18}, 385 (2002).

\bibitem{Wootters2}V. Coffman, J. Kundu, and W. K. Wootters, \pra {\bf 61}, 052306
(2000).

\bibitem{Lidar}S. Bandyopadhyay and D. A. Lidar, \pra {\bf 72}, 042339
(2005).

\bibitem{Zhou} D. L. Zhou, \prl \textbf{101}, 180505 (2008).

\bibitem{Molmer2}A. S\o rensen and K. M\o lmer, \prl {\bf 83}, 2274
(1999).

\bibitem{WangMolmer2}
X. Wang, A. S. S\o rensen, and K. M\o lmer, \pra {\bf 64}, 053815
(2001).

\end{thebibliography}
\end{document}